\documentclass{aa}
\usepackage{graphicx}

\def\mb#1{\setbox0=\hbox{$#1$}\kern-.025em\copy0\kern-\wd0
\kern-0.05em\copy0\kern-\wd0\kern-.025em\raise.0233em\box0}

\begin{document}
   \title{Gravitational instability of slowly rotating isothermal spheres }

   \author{P.H. Chavanis}

\institute{ Laboratoire de Physique Quantique, Universit\'e Paul Sabatier, 118
route de Narbonne 31062 Toulouse, France\\
\email{chavanis@irsamc2.ups-tlse.fr}}

   \date{To be included later }

   \abstract{We discuss the statistical mechanics of rotating
   self-gravitating systems by allowing properly for the conservation
   of angular momentum. We study analytically the case of slowly
   rotating isothermal spheres by expanding the solutions of the
   Boltzmann-Poisson equation in a series of Legendre polynomials,
   adapting the procedure introduced by Chandrasekhar (1933) for
   distorted polytropes. We show how the classical spiral of
   Lynden-Bell \& Wood (1967) in the temperature-energy plane is
   deformed by rotation. We find that gravitational instability
   occurs sooner in the microcanonical ensemble and later in the
   canonical ensemble. According to standard turning point arguments,
   the onset of the collapse coincides with the minimum energy or
   minimum temperature state in the series of
   equilibria. Interestingly, it happens to be close to the point of
   maximum flattening. We determine analytically the
   generalization of the singular isothermal solution to the case of a
   slowly rotating configuration. We also consider slowly rotating
   configurations of the self-gravitating Fermi gas at non zero
   temperature.
  
   }

   \maketitle
%

\section{Introduction}
\label{sec_introduction}

Recently, the statistical mechanics of self-gravitating systems has
attracted considerable attention (Chavanis, Sommeria \& Robert 1996,
Chavanis \& Sommeria 1998, de Vega, Sanchez \& Combes 1998, Youngkins
\& Miller 2000, Follana \& Laliena 2000, Semelin, Sanchez \& de Vega 2001, 
Cerruti-Sola, Cipriani \& Pettini 2001, Ispolatov \& Cohen 2001,
Chavanis 2002a, Taruya \& Sakagami 2002, de Vega \& Sanchez 2002,
Huber \& Pfenniger 2002...). This topic was introduced in the $1960$s
by Antonov (1962) and Lynden-Bell \& Wood (1968) and further developed
by Hertel \& Thirring (1971), Horwitz \& Katz (1978), Katz (1978) and
Padmanabhan (1989) among others (see a complete list of references in
the review of  Padmanabhan, 1990). These authors
pointed out the particularity of self-gravitating systems to possess
negative specific heats. They showed that this strange property is
responsible for the inequivalence of statistical ensembles
(microcanonical/canonical) and the occurence of giant phase
transitions associated with gravitational collapse. For a long time,
these topics were only discussed in the astrophysical literature and
were considered as a curiosity (not to say a fallacy) from statistical
mechanicians. The situation is changing lately as these properties are
re-discovered for other physical systems with long-range interactions
which can be studied in the laboratory (see, e.g., Gross 2001). For
that reason, maybe, the statistical mechanics of self-gravitating
systems comes back to fashion with new perspectives.

From our point of view, the statistical mechanics of self-gravitating
systems is far from being completely understood and rests on
simplifying idealizations. The first idealization is to enclose the
system within a box so as to prevent evaporation. It is only under
this condition (or by introducing more realistic truncated models)
that a rigorous statistical mechanics of self-gravitating systems can
be carried out. Second, for most astrophysical systems, the relaxation
time by two-body encounters is much larger than the age of the
universe so that a more subtle, collisionless, relaxation must be
advocated to explain the structure of galaxies. This is the concept of
{\it violent relaxation} formalized by Lynden-Bell in 1967. Then, it
is implicitly {\it assumed} that the relaxation towards statistical
equilibrium proceeds to completion, which is not necessarily the case
in reality. Indeed, it is possible that the relaxation stops before
the maximum entropy state is attained (see Lynden-Bell 1967, Tremaine,
H\'enon \& Lynden-Bell 1987, Chavanis et al. 1996). This problem of
{\it incomplete relaxation} must be approached with extensive
numerical simulations. Of course, this program was started long ago
(e.g., van Albada 1982) but only recently are $N$-body simulations
carefully compared with the predictions of the statistical mechanics
approach (Cerriti-Sola et al. 2001, Huber \& Pfeninger 2002) with
variable success.

The statistical equilibrium of a non-rotating classical gas enclosed
within a box was first investigated by Antonov (1962). He worked in
the microcanonical ensemble and found that thermodynamical equilibrium
exists only above a critical energy $E_{min}=-0.335GM^{2}/R$. Below
that energy, the system is expected to collapse and overheat; this is
the so-called ``gravothermal catastrophe'' (Lynden-Bell \& Wood
1968). An isothermal collapse also occurs below a critical temperature
$T_{min}={GMm\over 2.52 R}$ in the canonical ensemble. This
thermodynamical instability is closely related to the dynamical Jeans
instability (Semelin et al. 2001, Chavanis 2002a). In a recent series
of papers, we considered some extensions of the Antonov problem to the
case of self-gravitating fermions (Chavanis \& Sommeria 1998, Chavanis
2002c), special and general relativity (Chavanis 2002b) and confined
polytropes (Chavanis 2002d). These studies complete previous works on
the subject. We also introduced a simple dynamical model of Brownian
particles in gravitational interaction (Chavanis, Rosier
\& Sire 2002). By introducing by hands a friction and a noise, we
force the system to increase entropy continuously, thereby avoiding
the problem of incomplete relaxation. This model can be used to test
precisely the ideas of equilibrium statistical mechanics
(inequivalence of ensembles, phase transitions, gravitational
instabilities...) and is sufficiently simple to allow for a thorough
analytical investigation of the collapse regime when an equilibrium
state does not exist.

In all these studies, the system is assumed to be non-rotating so that
the conservation of angular momentum is trivially satisfied. The
object of the present paper is to extend the statistical mechanical
approach to the case of  rotating self-gravitating
systems. This problem has been considered previously by Lagoute \&
Longaretti (1996), Laliena (1999), Lynden-Bell (2000) and Fliegans \&
Gross (2002) with different types of models. Clearly, the most
interesting situation is the case of rapidly rotating systems since a
wide variety of structures can emerge as maximum entropy states
(Votyakov et al. 2002).  However, we shall restrict ourselves in the
present paper to the case of slowly rotating systems for which the
problem can be tackled analytically. We shall adapt to the case of
isothermal spheres the classical procedure developed by Chandrasekhar
(1933) for distorted polytropes, i.e. we shall expand the solutions of
the Boltzmann-Poisson equation in terms of Legendre polynomials. A
similar procedure was performed by Lagoute \& Longaretti (1996) for
rotating globular clusters subject to tidal forces and described by an
extended Michie-King model. We believe that it is useful to consider
the case of a gas enclosed within a box so as to make a clear
connexion with the Antonov model when the rotation is set to zero. In
particular, we shall derive the expression of the thermodynamical
parameters for slowly rotating isothermal spheres and show how the
classical spiral of Lynden-Bell \& Wood (1968) in the $E-T$ plane is
modified by rotation. We shall show that rotation avances the onset of
gravothermal catastrophe in the microcanonical ensemble and delays the
isothermal collapse in the canonical ensemble. Using the turning point
criterion of Katz (1978), we argue that the series of equilibrium
becomes unstable at the point of minimum energy (in the microcanonical
ensemble) or at the point of minimum temperature (in the canonical
ensemble). Interestingly, these instabilities happen to be close, in
each ensemble, to the point of maximum flattening. We establish the
generalization of the singular isothermal solution to the case of a
slowly rotating configuration.  We also consider the case of slowly
rotating self-gravitating fermions. This system exhibits phase transitions 
between ``gaseous'' states with an almost uniform
distribution of matter and ``condensed'' states with a core-halo
structure. By cooling below a critical temperature, an almost
nonrotating gaseous medium can collapse into a rotating ``fermion
ball'' containing a large fraction of mass and angular momentum.

\section{Statistical mechanics of rotating self-gravitating systems}
\label{sec_smrot}

\subsection{The mean-field approach}
\label{sec_mf}

Consider a system of $N$ particles, each of mass $m$, interacting via Newtonian gravity. We allow the system to have a non vanishing angular momentum. Let $f({\bf r},{\bf v},t)$ denote the distribution function of the system, i.e. $f({\bf r},{\bf v},t)d^{3}{\bf r}d^{3}{\bf v}$ gives the mass of particles whose position and velocity are in the cell $({\bf r},{\bf v};{\bf r}+d^{3}{\bf r},{\bf v}+d^{3}{\bf v})$ at time $t$. The integral of $f$ over the velocity determines the spatial density 
\begin{equation}
\rho=\int fd^{3}{\bf v}.
\label{mf1}
\end{equation} 
The mass and angular momentum of the configuration are given by
\begin{equation}
M=\int \rho d^{3}{\bf r},
\label{mf2}
\end{equation} 
\begin{equation}
{\bf L}=\int f\ {\bf r}\times {\bf v} \ d^{3}{\bf r}d^{3}{\bf v}.
\label{mf3}
\end{equation} 
On the other hand, in the mean-field approximation, the energy can be expressed
as 
\begin{equation}
E={1\over 2}\int f v^{2}d^{3}{\bf r}d^{3}{\bf v}+{1\over 2}\int \rho\Phi d^{3}{\bf r}=K+W,
\label{mf4}
\end{equation}
where $K$ is the kinetic energy and $W$ the potential energy. The gravitational potential $\Phi$ is related to the star density by the Newton-Poisson equation
\begin{equation}
\Delta\Phi=4\pi G\rho.
\label{mf5}
\end{equation}

The equilibrium configuration of the system is determined by maximizing the Boltzmann entropy  
\begin{equation}
S=-k\int {f\over m}\ln {f\over m}d^{3}{\bf r}d^{3}{\bf v},
\label{mf6}
\end{equation}
while conserving mass, angular momentum and energy. We stress that for systems interacting via a long-range potential, like gravity, the mean-field approximation is {\it exact} so that this procedure is entirely rigorous and provides a simple approach to the problem.

Introducing Lagrange multipliers $\alpha$, $\beta$ and $-\beta{\mb\Omega}$ for each constraint, we find that the critical points of entropy are given by
\begin{equation}
f=A'e^{-\beta ({v^{2}\over 2}+\Phi-{\bf\Omega}\cdot({\bf r}\times{\bf v}))},
\label{mf7}
\end{equation} 
where 
\begin{equation}
\beta={m\over kT}.
\label{mf8}
\end{equation} 
We can rewrite the foregoing expression for $f$ in the more suggestive form
\begin{equation}
f=A'e^{-{1\over 2}\beta ({\bf v}-{\bf \Omega}\times{\bf r})^{2}}e^{-\beta(\Phi-{1\over 2}({\bf\Omega}\times {\bf r})^{2})}.
\label{mf9}
\end{equation} 
We note that the most probable distribution of a rotating
self-gravitating system is a Maxwell-Boltzmann law with a dispersion
the same at every point (isothermal distribution). Moreover, the most
probable form of rotation is a rigid rotation ${\bf\Omega}$.  By
integrating Eq. (\ref{mf9}) over the velocity, we find that the
density is given by the Boltzmann factor
\begin{equation}
\rho=A e^{-\beta(\Phi-{1\over 2}({\bf\Omega}\times {\bf r})^{2})}.
\label{mf10}
\end{equation}    
The quantity in parenthesis is the ``effective'' potential in the rotating frame
\begin{equation}
\Phi_{eff}=\Phi-{1\over 2}({\bf\Omega}\times {\bf r})^{2},
\label{mf10a}
\end{equation}   
accounting for inertial forces. The equilibrium configuration is obtained by solving the Boltzmann-Poisson system
\begin{equation}
\Delta\Phi=4\pi GA e^{-\beta(\Phi-{1\over 2}({\bf\Omega}\times {\bf r})^{2})}.
\label{mf11}
\end{equation} 
and relating the Lagrange multipliers $A$, $\beta$ and ${\bf \Omega}$ to the constraints $M$, $E$ and ${\bf L}$.  The configuration is thermodynamically stable if the second order variations of entropy are negative for any perturbation that satisfies the constraints on mass, energy and angular momentum. This corresponds to the microcanonical description. Alternatively, we could describe the system in the canonical ensemble assuming that $\beta$ and ${\bf \Omega}$ are given, instead of $E$ and ${\bf L}$. In that case, the equilibrium configuration is obtained by maximizing the free energy $J=S-\beta E-\beta {\bf \Omega}\cdot{\bf L}$ at fixed $M$, $\beta$ and ${\bf\Omega}$. Of course, the microcanonical and canonical ensembles yield the same critical points, i.e. the critical points of entropy at fixed mass, energy and angular momentum and the critical points of free energy at fixed mass, temperature and angular velocity coincide. Only the onset of instability, regarding the second order variations of $S$ or $J$, will differ from an ensemble to the other. In the non-rotating case, the thermodynamical stability analysis was performed by Padmanabhan (1989) in the microcanonical ensemble and by Chavanis (2002a) in the canonical ensemble, by solving an eigenvalue equation. The onset of instability can also be determined by the turning point criterion of Katz (1978) who extended the theory of Poincar\'e on linear series of equilibrium. 

\subsection{The rotating isothermal sphere}
\label{sec_rotiso}

To determine the structure of rotating isothermal spheres, we first introduce the function $\Psi=\beta(\Phi_{eff}-\Phi_{0})$, where $\Phi_{0}$ is the gravitational potential at $r=0$. Then, the density field can be written
\begin{equation}
\rho=\rho_{0}e^{-\Psi},
\label{rot1}
\end{equation} 
where $\rho_{0}$ is the central density. Introducing the notations $\xi=(4\pi G\beta\rho_{0})^{1/2}r$, $v=\Omega^{2}/2\pi G\rho_{0}$ and using spherical coordinates $r,\theta,\phi$, we can rewrite the Boltzmann-Poisson equation (\ref{mf11}) in the form
\begin{equation}
{1\over\xi^{2}}{\partial\over\partial\xi}\biggl (\xi^{2}{\partial\Psi\over\partial \xi}\biggr )+{1\over\xi^{2}}{\partial\over\partial\mu}\biggl ((1-\mu^{2}){\partial\Psi\over\partial \mu}\biggr )=e^{-\Psi}-v,
\label{rot2}
\end{equation} 
where $\mu=\cos\theta$ and we have neglected $\phi$ by assuming that the configuration is symmetric with respect to the axis ${\bf\Omega}=\Omega {\bf z}$. The true gravitational potential $\Phi$ is related to $\Psi$ by the relation
\begin{equation}
\beta\Phi=\Psi+{1\over 4}v\xi^{2}(1-\mu^{2})+\beta\Phi_{0}.
\label{rot3}
\end{equation} 
Eq. (\ref{rot2}) is the fundamental equation of the problem. For $v=0$ (no rotation), we recover the Emden equation
\begin{equation}
{1\over\xi^{2}}{d\over d\xi}\biggl (\xi^{2}{d\psi\over d\xi}\biggr )=e^{-\psi}.
\label{rot4}
\end{equation} 
with $\psi=\psi'=0$ at $\xi=0$ (Chandrasekhar 1942).

So far, we have made no approximation regarding the value of the angular velocity. We shall now consider the case of slowly rotating structures and let $v\rightarrow 0$. Assuming the following form for our solution
\begin{equation}
\Psi=\psi+v\Theta+v^{2}\Theta_{2}+...
\label{rot5}
\end{equation} 
and keeping terms only to first order in $v$, we find that $\Theta$ satisfies the equation
\begin{eqnarray}
{1\over\xi^{2}}{\partial\over\partial\xi}\biggl (\xi^{2}{\partial\Theta\over\partial \xi}\biggr )+{1\over\xi^{2}}{\partial\over\partial\mu}\biggl ((1-\mu^{2}){\partial\Theta\over\partial \mu}\biggr )=-\Theta e^{-\Psi}-1.\nonumber\\
\label{rot6}
\end{eqnarray} 
Now, following a procedure that dates back to Milne (1923) and
Chandrasekhar (1933), we shall assume for $\Theta$ the following form
\begin{equation}
\Theta=\phi_{0}(\xi)+\sum_{j=1}^{+\infty}A_{j}\phi_{j}(\xi)P_{j}(\mu),
\label{rot7}
\end{equation} 
where $P_{j}(\mu)$ are the Legendre polynomials satisfying the differential equation
\begin{equation}
{\partial\over\partial\mu}\biggl ((1-\mu^{2}){\partial P_{j}\over\partial \mu}\biggr )+j(j+1)P_{j}=0.
\label{rot8}
\end{equation} 
Substituting for $\Theta$ from Eq. (\ref{rot7}) in Eq. (\ref{rot6}) and equating coefficients of $P_{j}$, we get
\begin{equation}
{1\over\xi^{2}}{d\over d\xi}\biggl (\xi^{2}{d\phi_{0}\over d\xi}\biggr )=-e^{-\psi}\phi_{0}-1,
\label{rot9}
\end{equation}
\begin{eqnarray}
{1\over\xi^{2}}{d\over d\xi}\biggl (\xi^{2}{d\phi_{j}\over d\xi}\biggr )=\Biggl ({j(j+1)\over\xi^{2}}-e^{-\psi}\biggr )\phi_{j}, \nonumber\\
(j=1,2,...)
\label{rot10}
\end{eqnarray}
with $\phi_{j}=\phi_{j}'=0$ at $\xi=0$.

So far, the $A_{j}$ are arbitrary. They will be determined by requiring that the gravitational potential and its radial derivative are continuous across the sphere at $r=R$. Now, outside the sphere the potential is given by the Laplace equation
\begin{equation}
\Delta\Phi_{ext}=0.
\label{rot11}
\end{equation}
The physically acceptable solution of this equation tending to zero at infinity can be written as
\begin{equation}
\beta\Phi_{ext}={B_{0}\over\xi}+v\sum_{j=1}^{+\infty}{B_{j}\over\xi^{j+1}}P_{j}(\mu)
\label{rot12}
\end{equation} 
On, the other hand, according to Eq. (\ref{rot3}), we have inside the spherical box
\begin{equation}
\beta\Phi_{int}=\Psi+{1\over 6}v\xi^{2}(1-P_{2}(\mu))+\beta\Phi_{0},
\label{rot13}
\end{equation}
where $\Psi(\xi,\mu)$ is given by Eqs. (\ref{rot5})(\ref{rot7}) and use has been made of the identity $1-\mu^{2}={2\over 3}(1-P_{2}(\mu))$. Let us denote by $\alpha=(4\pi G\beta\rho_{0})^{1/2}R$ the value of $\xi$ at $r=R$. Comparing the inner and the external potentials at $\xi=\alpha$, and also their derivative, we find that $A_{j}=B_{j}=0$ if $j\neq 2$ and, for $j=2$, 
\begin{equation}
{B_{2}\over\alpha^{3}}=A_{2}\phi_{2}(\alpha)-{1\over 6}\alpha^{2},
\label{rot14}
\end{equation}
\begin{equation}
-{3B_{2}\over\alpha^{4}}=A_{2}\phi_{2}'(\alpha)-{1\over 3}\alpha.
\label{rot15}
\end{equation}
Solving for $A_{2}$, we get
\begin{equation}
A_{2}={5\over 6}{\alpha^{2}\over 3\phi_{2}(\alpha)+\alpha\phi_{2}'(\alpha)}.
\label{rot16}
\end{equation}
Therefore, the solution of the Boltzmann-Poisson equation (\ref{rot2}) to first order in $v$ is given by
\begin{eqnarray}
\Psi=\psi(\xi)+v\biggl \lbrace \phi_{0}(\xi)+{5\over 6}{\alpha^{2}\over 3\phi_{2}(\alpha)+\alpha\phi_{2}'(\alpha)}\phi_{2}(\xi)P_{2}(\mu)\biggr\rbrace,\nonumber\\
\label{rot17}
\end{eqnarray}
with
\begin{equation}
{1\over\xi^{2}}{d\over d\xi}\biggl (\xi^{2}{d\phi_{0}\over d\xi}\biggr )=-e^{-\psi}\phi_{0}-1,
\label{rot18}
\end{equation}
\begin{eqnarray}
{1\over\xi^{2}}{d\over d\xi}\biggl (\xi^{2}{d\phi_{2}\over d\xi}\biggr )=\biggl ({6\over\xi^{2}}-e^{-\psi}\biggr )\phi_{2},
\label{rot19}
\end{eqnarray}
with $\phi_{j}=\phi_{j}'=0$ at $\xi=0$.

\subsection{The slowly rotating singular isothermal sphere}
\label{sec_sing}

For $\xi\rightarrow +\infty$, the solution of the Emden equation (\ref{rot4}) behaves like (Chandrasekhar 1942)
\begin{equation}
e^{-\psi}\sim {2\over\xi^{2}}.
\label{sing1}
\end{equation}
Substituting this asymptotic behaviour in Eq. (\ref{rot18}) and making the change of variables $\xi=e^{t}$, we obtain
\begin{equation}
{d^{2}\phi_{0}\over dt^{2}}+{d\phi_{0}\over dt}+2\phi_{0}=-e^{2t}.
\label{sing2}
\end{equation} 
The solution of this equation is readily found. After returning to original variables, we get for large $\xi$'s:
\begin{equation}
\phi_{0}=-{1\over 8}\xi^{2}+{A\over\xi^{1/2}}\cos\biggl ({\sqrt{7}\over 2}\ln\xi+\delta\biggr ).
\label{sing3}
\end{equation}
Following the same procedure for the function $\phi_{2}$, we have
\begin{equation}
{d^{2}\phi_{2}\over dt^{2}}+{d\phi_{2}\over dt}-4\phi_{2}=0,
\label{sing4}
\end{equation} 
 leading to
\begin{equation}
\phi_{2}={A'\over\xi^{1/2}}\cosh\biggl ({\sqrt{17}\over 2}\ln\xi+\delta'\biggr ).
\label{sing5}
\end{equation}

Keeping only the leading term in Eqs. (\ref{sing3})(\ref{sing5}) in the limit $\xi\rightarrow +\infty$, we find that
\begin{equation}
\phi_{0}\sim -{1\over 8}\xi^{2}; \qquad \phi_{2}\sim K\xi^{\gamma}, \qquad\gamma={\sqrt{17}\over 2}-{1\over 2}.
\label{sing6}
\end{equation}
Substituting these results in Eq. (\ref{rot17}) and returning to dimensional variables, we finally obtain
\begin{eqnarray}
\rho={1\over 2\pi G\beta r^{2}}+{\Omega^{2}\over 8\pi G}\biggl\lbrace 1-{20\over 3(3+\gamma)}\biggl ({R\over r}\biggr )^{2-\gamma}P_{2}(\mu)\biggr\rbrace,
\label{sing7}
\end{eqnarray}
which is the expression of the slowly rotating singular isothermal sphere.

\section{The thermodynamical parameters}
\label{sec_thermo}

\subsection{The mass}
\label{sec_mass}

In spherical coordinates, the mass is given by
\begin{equation}
M=2\pi\int_{-1}^{+1}\int_{0}^{R}\rho r^{2}drd\mu.
\label{mass1}
\end{equation}
Introducing the dimensionless variables previously defined, we obtain, using Eq. (\ref{rot1}),
\begin{equation}
\eta\equiv {\beta GM\over R}={1\over 2\alpha}\int_{-1}^{+1}\int_{0}^{\alpha}e^{-\Psi}\xi^{2}d\xi d\mu.
\label{mass2}
\end{equation}
Substituting for $\Psi$ from Eq. (\ref{rot17}) in Eq. (\ref{mass2}) and recalling that $\int_{-1}^{+1}P_{j}(\mu)d\mu=0$, we obtain to first order in $v$:
\begin{equation}
\eta={1\over \alpha}\int_{0}^{\alpha}e^{-\psi}(1-v\phi_{0}(\xi))\xi^{2}d\xi.
\label{mass3}
\end{equation}
Now,
\begin{equation}
\int_{0}^{\alpha}e^{-\psi}\xi^{2}d\xi=\int_{0}^{\alpha}d\xi{d\over d\xi}\biggl (\xi^{2}{d\psi\over d\xi}\biggr )=\alpha^{2}\psi'(\alpha),
\label{mass4}
\end{equation}
\begin{eqnarray}
\int_{0}^{\alpha}e^{-\psi}\phi_{0}\xi^{2}d\xi=-\int_{0}^{\alpha}d\xi \biggl\lbrace \xi^{2}+{d\over d\xi}\biggl (\xi^{2}{d\phi_{0}\over d\xi}\biggr )\biggr\rbrace\nonumber\\
=-{\alpha^{3}\over 3}-\alpha^{2}\phi_{0}'(\alpha).
\label{mass5}
\end{eqnarray}
Therefore, the normalized inverse temperature $\eta$ is expressed in terms of $\alpha$ and $v$ by the relation
\begin{equation}
\eta=\alpha\psi'(\alpha)+v\biggl \lbrack {\alpha^{2}\over 3}+\alpha\phi_{0}'(\alpha)\biggr\rbrack.
\label{mass6}
\end{equation}
Expressing the central density in terms of $\alpha$, we find that
\begin{equation}
v={2\eta\over\alpha^{2}}\omega^{2},\qquad {\rm with}\qquad  \omega=\Omega\biggl ({R^{3}\over GM}\biggr )^{1/2}.
\label{mass7}
\end{equation}
Therefore, the foregoing expression for $\eta$ can be rewritten to order $\omega^{2}$:
\begin{equation}
\eta=\alpha\psi'(\alpha)\biggl\lbrace 1+2\omega^{2}\biggl\lbrack {1\over 3}+{1\over\alpha}\phi_{0}'(\alpha)\biggr\rbrack\biggr\rbrace.
\label{mass8}
\end{equation}
The validity of our perturbative approach requires that 
\begin{equation}
\epsilon\equiv 2\omega^{2}\biggl\lbrack {1\over 3}+{1\over\alpha}\phi_{0}'(\alpha)\biggr\rbrack\ll 1.
\label{mass9}
\end{equation}

\subsection{The angular momentum}
\label{sec_angular}

Using the Maxwell-Boltzmann distribution (\ref{mf9}), we can rewrite the angular momentum (\ref{mf3}) in the form
\begin{equation}
{\bf L}=\int\rho\ {\bf r}\times {\bf u}\ d^{3}{\bf r},\qquad {\rm with}\qquad {\bf u}={\bf\Omega}\times{\bf r}.
\label{angular1}
\end{equation}
After straightforward manipulations, we find that the angular momentum is related to the angular velocity by
\begin{equation}
L_{i}=I_{ij}\Omega_{j},
\label{angular2}
\end{equation} 
where $I_{ij}=\int \rho (r^{2}\delta_{ij}-r_{i}r_{j})d^{3}{\bf r}$ is the moment of inertia tensor. If the density depends only on $r$ and $\theta$, we have ${\bf L}=I{\bf \Omega}$ with 
\begin{equation}
I=2\pi\int_{-1}^{+1}\int_{0}^{R} \rho r^{4}(1-\mu^{2})drd\mu.
\label{angular3}
\end{equation} 
For our purpose, it is sufficient to determine the angular momentum
to first order in $\Omega$. Therefore, we need just the expression of
the moment of inertia $I$ for a non-rotating isothermal
sphere. Introducing the dimensionless variables previously defined,
and using the expression (\ref{mass8}) for $\eta$, we get
\begin{equation}
{\cal I}\equiv {I\over MR^{2}}={2\over 3\alpha^{4}\psi'(\alpha)}\int_{0}^{\alpha} e^{-\psi} \xi^{4}d\xi.
\label{angular4}
\end{equation}
Then,
\begin{eqnarray}
\lambda={\cal I}(\alpha)\omega, \quad {\rm with}\quad \lambda={L\over \sqrt{GM^{3}R}}.
\label{angular5}
\end{eqnarray} 

\subsection{The energy}
\label{sec_energy}

Quite generally, the potential energy of a self-gravitating system can be written in the form (see, e.g., Binney \& Tremaine 1987) 
\begin{equation}
W=-\int\rho\ {\bf r}\cdot \nabla\Phi\ d^{3}{\bf r}.
\label{energy1}
\end{equation} 
It can be readily verified that the condition $f=f({w^{2}\over 2}+\Phi_{eff})$, where ${\bf w}={\bf v}-{\bf u}$ and $\Phi_{eff}$ is the effective potential (\ref{mf10a}), is  equivalent to the condition of hydrostatic equilibrium in the rotating frame
\begin{equation}
\nabla p=-\rho\nabla\Phi-\rho{\bf\Omega}\times({\bf\Omega}\times{\bf r}),
\label{energy2}
\end{equation}
where
\begin{equation}
p={1\over 3}\int fw^{2}d^{3}{\bf w},
\label{energy2bis}
\end{equation}
is the local pressure. For an isothermal gas $p={k\over m}\rho T$.  Inserting Eq. (\ref{energy2}) in Eq. (\ref{energy1}), we obtain after straightforward calculations
\begin{equation}
W+3\int p \ d^{3}{\bf r}+{\bf L}\cdot{\bf\Omega}-\oint p\ {\bf r}\cdot d{\bf S}=0,
\label{energy3}
\end{equation}
where $d{\bf S}$ is a surface element normal to the spherical box.  On
the other hand, we can write the kinetic energy
in the form
\begin{equation}
K=K_{rot}+K_{th},
\label{energy4}
\end{equation} 
where 
\begin{equation}
K_{rot}={1\over 2}\int\rho u^{2}d^{3}{\bf r}= 
{1\over 2}{\bf L}\cdot{\bf\Omega},
\label{energy4bis}
\end{equation} 
is the rotational energy and 
\begin{equation}
K_{th}={1\over 2}\int f{w^{2}}d^{3}{\bf r}d^{3}{\bf w}={3\over 2}\int p d^{3}{\bf r}, 
\label{energy4tris}
\end{equation} 
is the thermal energy. For an isothermal gas $K_{th}={3\over 2}NkT$.
Therefore, Eq. (\ref{energy3}) becomes
\begin{equation}
W+2K=\oint p\ {\bf r}\cdot d{\bf S},
\label{energy5}
\end{equation}
which is the Virial theorem for a self-gravitating gas enclosed within a box.
For the total energy $E=K+W$, we have
\begin{equation}
E=-K+\oint p\ {\bf r}\cdot d{\bf S}.
\label{energy6}
\end{equation} 
This expression is valid even if the system is not axisymmetric. 
If, now, the density depends only on $r$ and $\theta$, and if the gas is isothermal, it is possible to combine the formulae derived in Sect. \ref{sec_rotiso} to obtain the expression 
\begin{eqnarray}
\Lambda\equiv -{ER\over GM^{2}}={3\over 2\eta}+{1\over 2}\lambda\omega-{\alpha^{2}\over \eta^{2}}e^{-\psi(\alpha)}( 1-v\phi_{0}(\alpha)),
\label{energy7}
\end{eqnarray} 
where $\eta$ and $v$ can be expressed in terms of $\alpha$ and $\omega$ by Eqs. (\ref{mass8})(\ref{mass7}). We also recall that our theory is valid to order $\omega^{2}$. For $\omega=0$, Eqs. (\ref{mass8})(\ref{energy7}) reduce to the equations of state obtained by Lynden-Bell \& Wood (1968) for non-rotating isothermal spheres.

\subsection{The entropy}
\label{sec_entropy}

Using Eqs. (\ref{mf9})(\ref{mf10}) and (\ref{rot1}), the distribution function of a rotating  isothermal gas can be written
\begin{eqnarray}
f=\biggl ({\beta\over 2\pi}\biggr )^{3/2}e^{-\beta {w^{2}\over 2}}\rho_{0}e^{-\Psi}.
\label{entropy1}
\end{eqnarray} 
Substituting this expression in Eq. (\ref{mf6}) and expressing $\rho_{0}$ in terms of $\alpha$, we get 
\begin{eqnarray}
{mS\over k}=-{1\over 2}M\ln\beta-2M\ln\alpha+\int\rho\Psi d^{3}{\bf r}.
\label{entropy2}
\end{eqnarray} 
Throughout this paper, we shall not write the constant terms
(depending on the fixed parameters $M$ and $R$) which enter in the
expression of the entropy. Therefore, a term $\beta K_{th}={3\over
2}M$ has been ignored in Eq. (\ref{entropy2}). Using the definition of
$\Psi$, the last integral can be rewritten
\begin{eqnarray}
\int\rho\Psi d^{3}{\bf r}=2\beta W-\beta K_{rot}-\beta M\Phi_{0}.
\label{entropy3}
\end{eqnarray}
Using $W=E-K_{rot}-K_{th}$, we obtain 
\begin{eqnarray}
{mS\over k}=-{1\over 2}M\ln\beta-{3\over 2}\beta \ {\bf L}\cdot {\bf\Omega}\nonumber\\
-2M\ln\alpha
+2\beta E-\beta M\Phi_{0}.
\label{entropy4}
\end{eqnarray}
Introducing the dimensionless parameters previously defined, we get
\begin{eqnarray}
{S\over Nk}=-{1\over 2}\ln\eta-{3\over 2}\eta\lambda\omega-2\ln\alpha-2\eta \Lambda-\beta \Phi_{0}.
\label{entropy5}
\end{eqnarray}
We now need to determine the central potential $\Phi_{0}$. The condition that $\Phi$ and $\partial\Phi/\partial\xi$ are continuous at $\xi=\alpha$ implies for $j=0$:
\begin{eqnarray}
{B_{0}\over \alpha}=\psi(\alpha)+v\phi_{0}(\alpha)+{1\over 6}v\alpha^{2}+\beta\Phi_{0},
\label{entropy6}
\end{eqnarray}
\begin{eqnarray}
-{B_{0}\over \alpha^{2}}=\psi'(\alpha)+v\phi_{0}'(\alpha)+{1\over 3}v\alpha.
\label{entropy7}
\end{eqnarray}
Comparing Eq. (\ref{entropy7}) with Eq. (\ref{mass6}), we see that $-B_{0}/\alpha=\eta$. Inserting this result in Eq. (\ref{entropy6}), we obtain
\begin{eqnarray}
-\beta\Phi_{0}=\eta+\psi(\alpha)+v\biggl\lbrack {\alpha^{2}\over 6}+\phi_{0}(\alpha)\biggr\rbrack.
\label{entropy8}
\end{eqnarray}
With this new relation, the entropy (\ref{entropy5}) becomes
\begin{eqnarray}
{S\over Nk}=-{1\over 2}\ln\eta-2\ln\alpha+\psi(\alpha)+\eta-2\Lambda\eta\nonumber\\
-{3\over 2}\eta\lambda\omega+v\biggl\lbrack {\alpha^{2}\over 6}+\phi_{0}(\alpha)\biggr\rbrack. 
\label{entropy9}
\end{eqnarray}

\subsection{The flattening function}
\label{sec_flat}

The value of the potential at $\xi$ for a non-rotating configuration is $\psi(\xi)$. For a rotating configuration, the equation of the surface with the same value of the potential is given, to first order in $v$, by
\begin{eqnarray}
\xi'=\xi-{v\over \psi'(\xi)}\biggl\lbrace\phi_{0}(\xi)+{5\over 6}{\alpha^{2}\over 3\phi_{2}(\alpha)+\alpha\phi_{2}'(\alpha)}\phi_{2}(\xi)P_{2}(\mu)\biggr\rbrace.\nonumber\\
\label{flat1}
\end{eqnarray}
If $a$ denotes the largest radius of the isodensity surface (at the equator $\mu=0$) and $b$ the smallest radius (at the pole $\mu=1$), and if we define the flattening by $f=1-b/a$ (see, e.g., Lagoute \& Longaretti 1996), we get
\begin{eqnarray}
f(\xi)={5\over 4}{v\over \xi\psi'(\xi)}{\alpha^{2}\over 3\phi_{2}(\alpha)+\alpha\phi_{2}'(\alpha)}\phi_{2}(\xi).
\label{flat2}
\end{eqnarray}
Considering the limit $\xi\rightarrow +\infty$ and returning to dimensional  variables, we find that the flattening function of the rotating  singular isothermal sphere is
\begin{eqnarray}
f(r)={5\over 4}{\beta\Omega^{2}R^{2}\over 3+\gamma}\biggl ({r\over R}\biggr )^{\gamma},
\label{flat3}
\end{eqnarray}
a result which can also be derived directly from Eq. (\ref{sing7}). 

Coming back to Eq. (\ref{flat2}), we find that the flattening behaves with the distance (for a given value of $\alpha$) as
\begin{eqnarray}
F(\xi)=-{\phi_{2}(\xi)\over \xi\psi'(\xi)}.
\label{flat4}
\end{eqnarray}
On the other hand, the flattening at the edge of the configuration 
($\xi=\alpha$) is given by
\begin{eqnarray}
f(\alpha)={5\over 2}\omega^{2}{\phi_{2}(\alpha)\over 3\phi_{2}(\alpha)+\alpha\phi_{2}'(\alpha)},
\label{flat5}
\end{eqnarray}
where we have used Eqs. (\ref{mass7})(\ref{mass8}) to eliminate the variable $v$ in profit of $\omega$. In the canonical ensemble (fixed $\omega$), this function depends on $\alpha$ as
\begin{eqnarray}
F_{CE}(\alpha)={\phi_{2}(\alpha)\over 3\phi_{2}(\alpha)+\alpha\phi_{2}'(\alpha)},
\label{flat6}
\end{eqnarray}
In the microcanonical ensemble (fixed $\lambda$), we must express the angular velocity in terms of the angular momentum using the relation (\ref{angular5}). In that case, the flattening at the edge of the configuration depends on $\alpha$  as 
\begin{eqnarray}
F_{MCE}(\alpha)={1\over {\cal I}(\alpha)^{2}}F_{CE}(\alpha).
\label{flat7}
\end{eqnarray}
We shall come back to these results in the following section.

\section{Equilibrium phase diagram}
\label{sec_epd}

\subsection{Microcanonical ensemble}
\label{sec_mce}

\begin{figure}
\centering
\includegraphics[width=8.8cm]{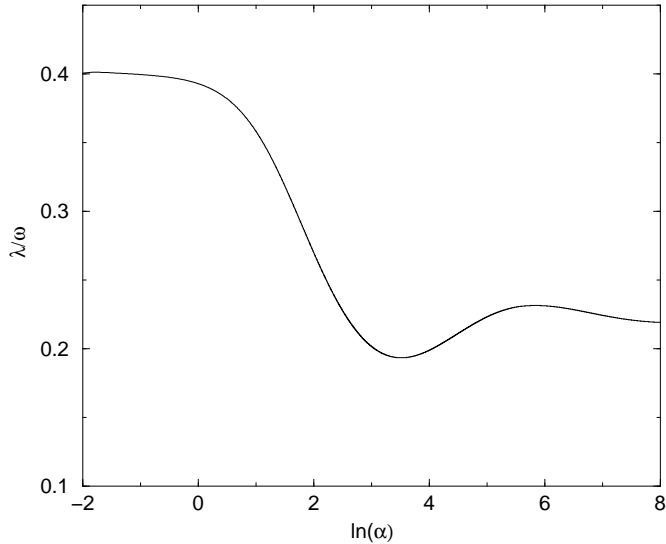}
\caption{Moment of inertia for a non-rotating isothermal sphere along the series of equilibria (parametrized by $\alpha$).}
\label{IP}
\end{figure}

\begin{figure}
\centering
\includegraphics[width=8.8cm]{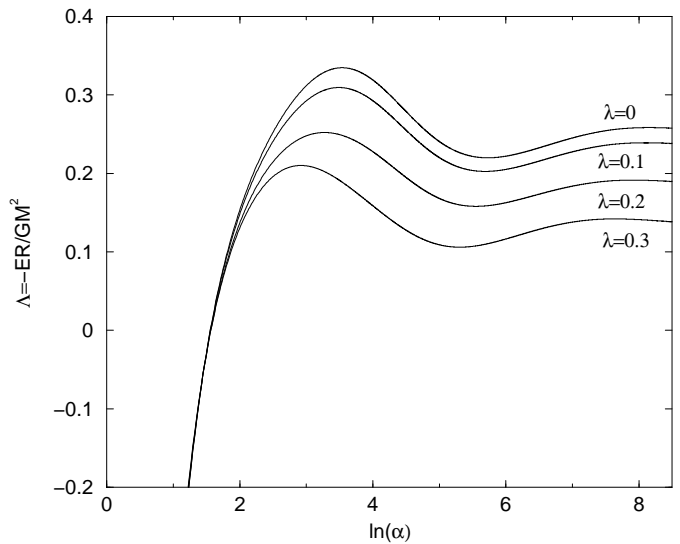}
\caption{Normalized energy of an isothermal sphere along the series of equilibria (parametrized by $\alpha$) for different values of the angular momentum $\lambda=0,0.1,0.2,0.3$. }
\label{alphaLambdaMICRO}
\end{figure}

\begin{figure}
\centering
\includegraphics[width=8.8cm]{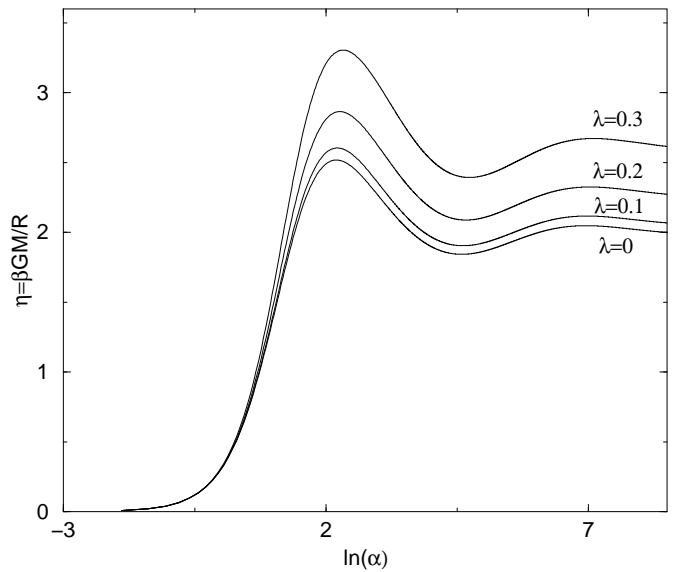}
\caption{Normalized inverse temperature of an isothermal sphere along the series of equilibria (parametrized by $\alpha$) for different values of the angular momentum $\lambda=0,0.1,0.2,0.3$. }
\label{alphaetaMICRO}
\end{figure}

\begin{figure}
\centering
\includegraphics[width=7cm]{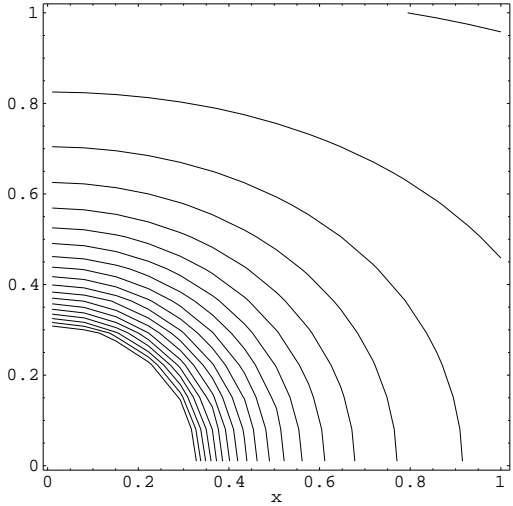}
\caption{Iso-density contours of the rotating singular isothermal sphere with an angular momentum $\lambda=0.17$.}
\label{courbeniveau}
\end{figure}

The microcanonical ensemble corresponds to isolated systems
characterized by their energy $\Lambda$ and their angular momentum
$\lambda$. In order to determine the
equilibrium phase diagram $\eta(\Lambda)$ for different values of
$\lambda$, we need to solve Eqs. (\ref{rot4})(\ref{rot18})(\ref{rot19}) 
numerically. Expanding the functions
$\psi$, $\phi_{0}$ and $\phi_{2}$ in Taylor series for $\xi\rightarrow
0$, we get
\begin{equation}
\psi={1\over 6}\xi^{2}-{1\over 120}\xi^{4}+...,
\label{epd1}
\end{equation}  
\begin{equation}
\phi_{0}=-{1\over 6}\xi^{2}+{1\over 120}\xi^{4}+...,
\label{epd2}
\end{equation} 
\begin{equation}
\phi_{2}=-\xi^{2}+{1\over 14}\xi^{4}+...
\label{epd3}
\end{equation}
so that $\psi''(0)=1/3$, $\phi_{0}''(0)=-1/3$ and $\phi_{2}''(0)=-2$. The integration can be continued numerically by a standard Runge-Kutta routine. In Figs. \ref{IP},\ref{alphaLambdaMICRO},\ref{alphaetaMICRO}, we plot the curves ${\cal I}(\alpha)$, $\Lambda(\alpha)$ and $\eta(\alpha)$ defined by Eqs. (\ref{angular4})(\ref{energy7})(\ref{mass8}). These curves exhibit damped oscillations and tend to the values
\begin{equation}
{\cal I}_{s}={2\over 9},\quad \eta_{s}=2+{27\over 4}\lambda^{2}, \quad \Lambda_{s}={1\over 4}-{63\over 32}\lambda^{2}.
\label{epd4}
\end{equation} 
as $\alpha\rightarrow +\infty$.  This asymptotic limit corresponds to the singular solution (\ref{sing7}). The iso-density contours of the rotating singular isothermal sphere are represented in Fig. \ref{courbeniveau} (for $\lambda=0.17$). 

\begin{figure}
\centering
\includegraphics[width=8.8cm]{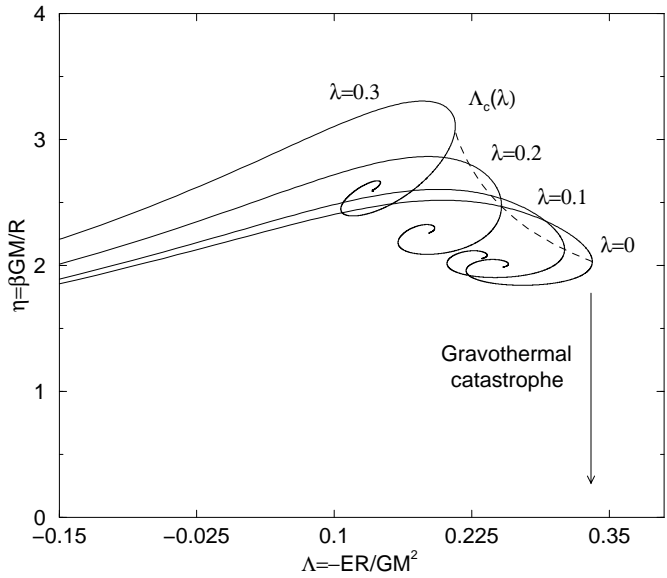}
\caption{Equilibrium phase diagram of isothermal spheres giving the inverse temperature $\eta$ as a function of minus the energy $\Lambda$ for different values of the angular momentum $\lambda$. The gravothermal instability at $\Lambda_{c}$ occurs sonner (i.e. for larger energies) when the system is rotating.}
\label{LambdaetaMICRO}
\end{figure}

In Fig. \ref{LambdaetaMICRO}, we have represented the curve $\eta(\Lambda)$. It has a classical spiral behaviour as noted by a number of authors in the non-rotating case. There is no equilibrium state (i.e., no critical point of entropy) above the value $\Lambda_{c}(\lambda)$. In that case, the system will collapse and overheat ({gravothermal catastrophe}). It is also at this point that the critical points of entropy become unstable (saddle points) in the series of equilibria (Katz 1978). We see that rotation tends to favour the instability, i.e., the gravothermal catastrophe occurs sooner than in the non-rotating case.

\begin{figure}
\centering
\includegraphics[width=8.8cm]{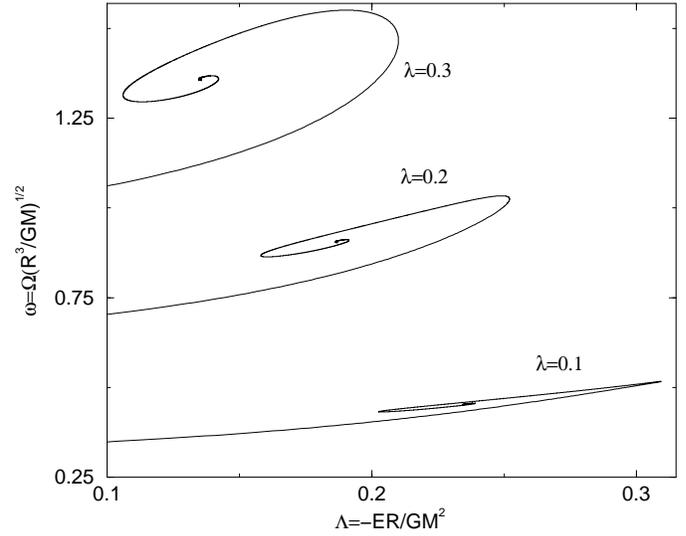}
\caption{Angular velocity $\omega$ of the system as a function of energy $\Lambda$ for 
different values of the angular momentum $\lambda$. }
\label{Lomega}
\end{figure}

In Fig. \ref{Lomega}, we have represented the angular velocity of the
system as a function of energy for different values of the angular
momentum. We observe that the curve has a spiral behavior similar to
the $\eta-\Lambda$ diagram, but reversed. We might expect that the
moment of inertia  decreases as the system becomes more and more
concentrated, resulting in an increase of angular velocity along
the series of equilibria. This is true for moderate density contrasts
(up to $\sim 709$), coinciding with the region of stability, but not
for larger density contrasts. Indeed, although the central density
tends to diverge, the mass contained in the core is low and does not
dominate the moment of inertia. Therefore, the moment of inertia and the
angular velocity have a non monotonous (in fact oscillatory) behaviour
with $\alpha$ and saturate to finite values ${\cal I}_{s}$ and
$\omega_{s}=\lambda/{\cal I}_{s}$ as $\alpha\rightarrow +\infty$.

\begin{figure}
\centering
\includegraphics[width=8.8cm]{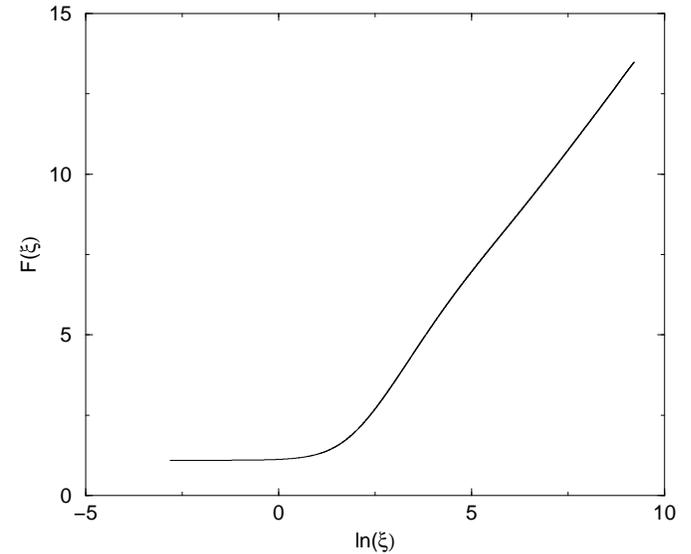}
\caption{Spatial dependance of the flattening function $F(\xi)$.}
\label{xiflatt}
\end{figure}

\begin{figure}
\centering
\includegraphics[width=8.8cm]{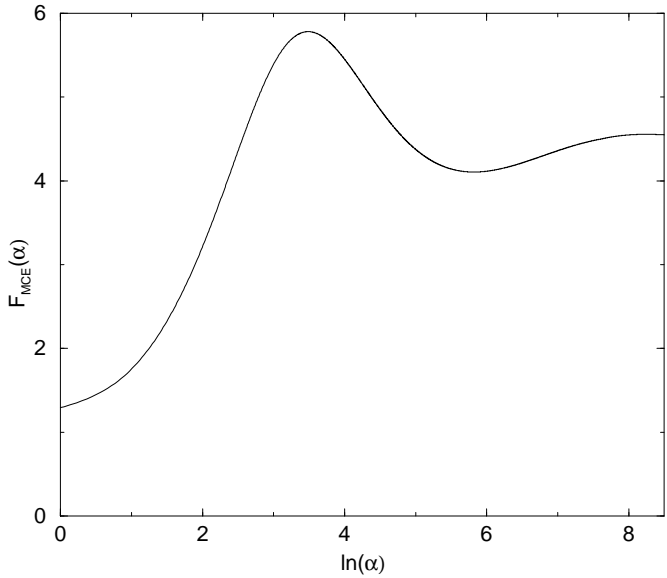}
\caption{Flattening at the edge of the configuration (per unit angular momentum squared) along the series of equilibria (microcanonical ensemble). The flattening is maximum for $\alpha=32.6$. This value lies typically in the region where the gravothermal catastrophe sets in.}
\label{alphaflattmicro}
\end{figure}

In Fig. \ref{xiflatt}, we plot the flattening function $F(\xi)$ defined by Eq. (\ref{flat4}). As expected, the flattening is is a monotonous function of the distance. For $\xi\rightarrow 0$, $F\rightarrow 3$ and for $\xi\rightarrow +\infty$, $F\sim \xi^{\gamma}$.  In Fig. \ref{alphaflattmicro}, we plot the flattening at the edge of the configuration $F_{MCE}(\alpha)$ as a function of $\alpha$ in the microcanonical ensemble (see Eq. (\ref{flat7})). We observe that the curve displays damped oscillations towards the asymptotic value ${1\over 3+\gamma}(9/2)^{2}$. In particular the flattening (by unit of $\lambda^{2}$) is maximum  for $\alpha=32.6...$. Interestingly, this value lies precisely in the range of values at which the gravothermal catastrophe sets in (compare with Fig. \ref{alphaLambdaMICRO}).

\subsection{Canonical ensemble}
\label{sec_ce}

\begin{figure}
\centering
\includegraphics[width=8.8cm]{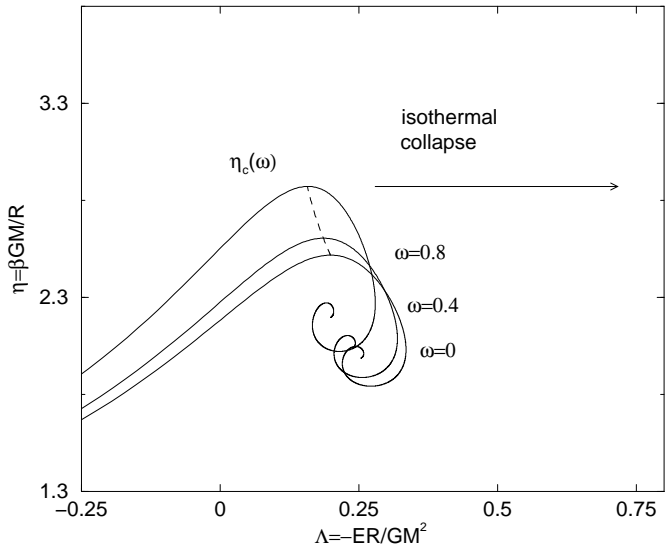}
\caption{Equilibrium phase diagram of isothermal spheres giving minus the energy $\Lambda$ as a function of the inverse temperature $\eta$ for different values of  angular velocity.  The isothermal collapse at $\eta_{c}(\omega)$ occurs later (i.e. for smaller temperatures) when the system is rotating.}
\label{LambdaetaCANO}
\end{figure}

The canonical ensemble is characterized by the specification of the
inverse temperature $\eta$ and the angular velocity $\omega$. In
Fig. \ref{LambdaetaCANO}, we have represented the curve
$\Lambda(\eta)$ for different values of $\omega$.  There is no
equilibrium state (i.e., no critical point of free energy) above the
value $\eta_{c}(\omega)$. In that case, the system will undergo an {
isothermal collapse}. It is also at this point that the solutions
become unstable (saddle points of free energy) in the series of
equilibria (Katz 1978). We see that rotation tends to delay the
instability, i.e., the isothermal collapse occurs later than in the
non-rotating case.

\begin{figure}
\centering
\includegraphics[width=8.8cm]{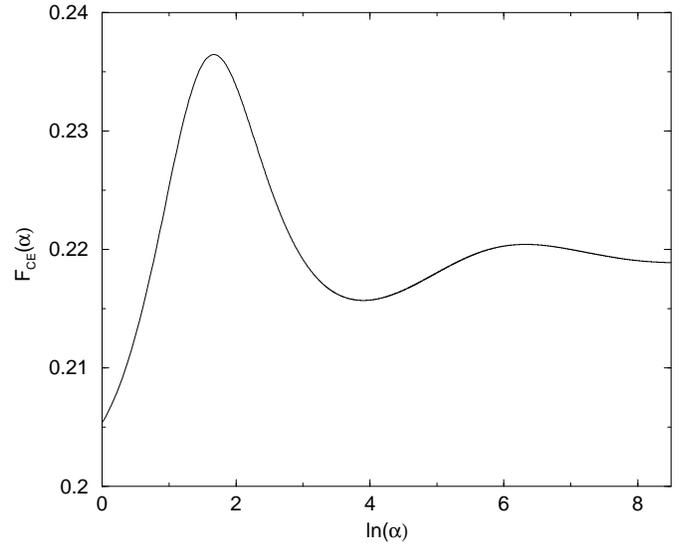}
\caption{Flattening at the edge of the configuration (per unit of angular velocity squared) along the series of equilibria (canonical ensemble). The flattening is maximum for $\alpha=5.4$. This value lies typically in the region where the isothermal collapse sets in. }
\label{alphaflattcano}
\end{figure}

\begin{figure}
\centering
\includegraphics[width=8.8cm]{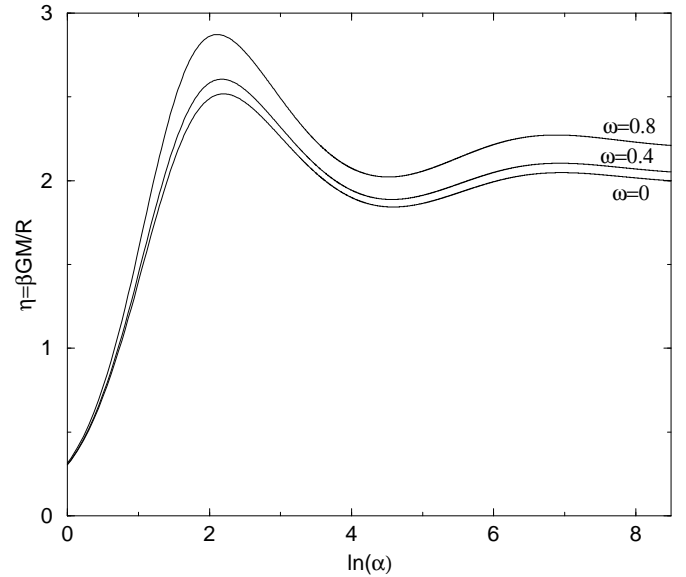}
\caption{Normalized inverse temperature of an isothermal sphere along the series of equilibria (parametrized by $\alpha$) for different values of angular velocity $\omega=0,0.4,0.8$. The series of equilibria becomes unstable after the first maximum.}
\label{alphaetaCANO}
\end{figure}

In Fig. \ref{alphaflattcano}, we plot the flattening at the edge of the configuration $F_{CE}(\alpha)$ as a function of $\alpha$ in the canonical ensemble. The curve displays damped oscillations towards the value ${1\over 3+\gamma}$. The flattening (by unit of $\omega^{2}$) is maximum for $\alpha=5.4...$. Interestingly, this value is close to the typical values at which the isothermal collapse sets in (compare with Fig. \ref{alphaetaCANO} giving the inverse temperature along the series of equilibria).

\section{Rotating self-gravitating fermions}
\label{sec_fermions}

The previous results can be easily generalized to the case of self-gravitating fermions. This extension is relatively straightforward and we shall just give the main steps of the calculations. The thermodynamical parameters of a non-rotating Fermi gas at finite temperature have been calculated by Chavanis \& Sommeria (1998) and we shall adopt a similar presentation. For a rotating configuration, the Fermi-Dirac distribution can be written
\begin{equation}
f={\eta_{0}\over 1+\lambda e^{\beta\Phi_{eff}}e^{\beta {w^{2}\over 2}}},
\label{fermions1}
\end{equation}
where $\eta_{0}$ is the maximum allowable value of the distribution function and $\lambda>0$ a strictly positive parameter (inverse of the fugacity) insuring that $f\le\eta_{0}$ (the other quantities have been defined previously). For quantum particles with spin $s$, $\eta_{0}=(2s+1)m^{4}/(2\pi \hbar)^{3}$. The distribution function (\ref{fermions1}) can be obtained by maximizing the Fermi-Dirac entropy
\begin{equation}
S=-\int\biggl\lbrace {f\over\eta_{0}}\ln {f\over\eta_{0}}+\biggl (1-{f\over\eta_{0}}\biggr )\ln\biggl (1-{f\over\eta_{0}}\biggr )\biggr\rbrace d^{3}{\bf r}d^{3}{\bf v},\nonumber\\ 
\label{fermions1bis}
\end{equation}
at fixed mass, energy and angular momentum. This form of entropy also
occurs in the context of violent relaxation for collisionless
self-gravitating systems (Lynden-Bell 1967, Chavanis, Robert \&
Sommeria 1996, Chavanis \& Sommeria 1998).  Defining
$\Psi=\beta(\Phi_{eff}-\Phi_{0})$ and $k=\lambda e^{\beta\Phi_{0}}$,
the spatial density can be written
\begin{equation}
\rho=\rho_{0}{I_{1/2}(ke^{\Psi})\over I_{1/2}(k)}, \quad {\rm with}\quad \rho_{0}={4\pi\sqrt{2}\eta_{0}\over\beta^{3/2}}I_{1/2}(k),
\label{fermions2}
\end{equation} 
where $I_{1/2}$ is the Fermi integral
\begin{eqnarray}
I_{n}(t)=\int_{0}^{+\infty}{x^{n}\over 1+te^{x}}dx
\label{fermions3}
\end{eqnarray} 
of order $n=1/2$. Substituting the relation (\ref{fermions2}) in the Poisson equation (\ref{mf5}) and introducing the dimensionless parameters 
$\xi=(4\pi G\beta\rho_{0})^{1/2}r$ and $v={\Omega^{2}\over 2\pi G\rho_{0}}$, we obtain
\begin{eqnarray}
{1\over\xi^{2}}{\partial\over\partial\xi}\biggl (\xi^{2}{\partial\Psi\over\partial \xi}\biggr )+{1\over\xi^{2}}{\partial\over\partial\mu}\biggl ((1-\mu^{2}){\partial\Psi\over\partial \mu}\biggr )={I_{1/2}(ke^{\Psi})\over I_{1/2}(k)}-v.\nonumber\\
\label{fermions4}
\end{eqnarray} 
We can also check that Eq. (\ref{rot3}) keeps the same form. For $v=0$, Eq. (\ref{fermions4}) reduces to 
\begin{eqnarray}
{1\over\xi^{2}}{d\over d\xi}\biggl (\xi^{2}{d\psi\over d\xi}\biggr )={I_{1/2}(ke^{\psi})\over I_{1/2}(k)},
\label{fermions5}
\end{eqnarray}
with $\psi=\psi'=0$ at $\xi=0$. For $k\rightarrow +\infty$ (classical limit), we can use the limiting form of the Fermi integral
\begin{eqnarray}
I_{n}(t)\sim {1\over t}\Gamma(n+1),\qquad (t\rightarrow +\infty)
\label{fermions5a}
\end{eqnarray}
and we recover the classical Emden equation (\ref{rot4}). For $k\rightarrow 0$ (completely degenerate limit), we have
\begin{eqnarray}
I_{n}(t)\sim {(-\ln t)^{n+1}\over n+1},\qquad (t\rightarrow 0)
\label{fermions5b}
\end{eqnarray}
and the differential equation (\ref{fermions5}) becomes equivalent to the Lane-Emden equation for a polytrope of index $3/2$.

We shall now consider the case of slowly rotating structures and let
$v\rightarrow 0$. Repeating the steps of Sec. \ref{sec_rotiso}, we
find that Eq. (\ref{rot17}) remains valid with the new functions
$\phi_{0}$ and $\phi_{2}$ defined by
\begin{eqnarray}
{1\over\xi^{2}}{d\over d\xi}\biggl (\xi^{2}{d\phi_{0}\over d\xi}\biggr )=-{1\over 2}{I_{-1/2}(ke^{\psi})\over I_{1/2}(k)}\phi_{0}-1,
\label{fermions6}
\end{eqnarray}
\begin{eqnarray}
{1\over\xi^{2}}{d\over d\xi}\biggl (\xi^{2}{d\phi_{2}\over d\xi}\biggr )=\biggl ({6\over\xi^{2}}-{1\over 2}{I_{-1/2}(ke^{\psi})\over I_{1/2}(k)}\biggr )\phi_{2},
\label{fermions7}
\end{eqnarray}
with $\phi_{j}=\phi'_{j}=0$ at $\xi=0$.  In arriving at (\ref{fermions6})-(\ref{fermions7}), we have used the identity
\begin{eqnarray}
I'_{n}(t)=-{n\over t}I_{n-1}(t), \qquad (n>0),
\label{fermions8}
\end{eqnarray}
for $n=1/2$, which can be easily established from Eq. (\ref{fermions3}). 

We now determine the thermodynamical parameters of a slowly rotating
self-gravitating Fermi gas. We can check that the inverse temperature
is still given by Eq. (\ref{mass8}) and that the relation
(\ref{mass7}) remains valid. On the other hand, eliminating the
central density between Eq. (\ref{fermions2}) and $\alpha=(4\pi
G\beta\rho_{0})^{1/2}R$, we find that
\begin{eqnarray}
\eta={\mu^{2}\over \alpha^{4}}I_{1/2}(k)^{2}, \qquad {\rm where}\quad \mu=\eta_{0}\sqrt{512\pi^{4}G^{3}MR^{3}},
\label{fermions9}
\end{eqnarray}
is the ``degeneracy parameter'' (Chavanis \& Sommeria 1998). Using
Eq. (\ref{mass8}), we obtain
\begin{eqnarray}
\mu^{2}I_{1/2}(k)^{2}=\alpha^{5}\psi'(\alpha)\biggl\lbrace 1+2\omega^{2}\biggl\lbrack {1\over 3}+{1\over\alpha}\phi_{0}'(\alpha)\biggr\rbrack\biggr\rbrace.
\label{fermions10}
\end{eqnarray}
The relation between the angular momentum and the angular velocity is still given by Eq. (\ref{angular5}) with the moment of inertia
\begin{eqnarray}
{\cal I}={2\over 3\alpha^{4}\psi'(\alpha)}\int_{0}^{\alpha}{I_{1/2}(ke^{\psi})\over I_{1/2}(k)}\xi^{4}d\xi.
\label{fermions11}
\end{eqnarray}
Using Eq. (\ref{energy6}), the energy can be put in the form
\begin{eqnarray}
\Lambda={\alpha^{7}\over\mu^{4}}\int_{0}^{\alpha}\biggl\lbrace {I_{3/2}(ke^{\psi})\over I_{1/2}(k)^{5}}-{3\over 2}v{I_{1/2}(ke^{\psi})\over I_{1/2}(k)^{5}}\phi_{0}(\xi)\biggr\rbrace\xi^{2}d\xi\nonumber\\
+{1\over 2}\lambda\omega-{2\over 3}{\alpha^{10}\over\mu^{4}}\biggl\lbrace {I_{3/2}(ke^{\psi(\alpha)})\over I_{1/2}(k)^{5}}-{3\over 2}v{I_{1/2}(ke^{\psi(\alpha)})\over I_{1/2}(k)^{5}}\phi_{0}(\alpha)\biggr\rbrace.\nonumber\\
\label{fermions12}
\end{eqnarray}
Finally, for the entropy we obtain (see Appendix \ref{sec_app2}) 
\begin{eqnarray}
{\eta_{0}S\over M}=\ln k+\eta+\psi(\alpha)-{7\over 3}\eta \Lambda -{4\over 3}\eta\lambda\omega+v\biggl\lbrack {\alpha^{6}\over 6}+\phi_{0}(\alpha)\biggr\rbrack\nonumber\\
-{2\over 9}{\alpha^{6}\over\mu^{2}}\biggl\lbrace {I_{3/2}(ke^{\psi(\alpha)})\over I_{1/2}(k)^{3}}-{3\over 2}v{I_{1/2}(ke^{\psi(\alpha)})\over I_{1/2}(k)^{3}}\phi_{0}(\alpha)\biggr\rbrace.\qquad\qquad
\label{fermions13}
\end{eqnarray}
For $v=0$, Eqs. (\ref{mass8}) (\ref{fermions10})(\ref{fermions12}) and (\ref{fermions13})  reduce to the equations of state obtained by Chavanis \& Sommeria (1998) for the non-rotating Fermi gas. For $k\rightarrow +\infty$ (non degenerate limit), we recover the equations derived in Sec. \ref{sec_thermo}.

The equilibrium phase diagram can be obtained in the following manner. For given $k$, $\mu$ and $\omega$, we can solve Eqs. (\ref{fermions5})
(\ref{fermions6}) and (\ref{fermions7}) until the value 
$\xi=\alpha$ for which the relation (\ref{fermions10}) is satisfied. Then, 
Eqs. (\ref{mass8}) and (\ref{fermions12}) determine the temperature and the energy of the configuration. If the angular momentum is fixed instead of the angular velocity, we must use Eq. (\ref{angular5}) with Eq. (\ref{fermions11}) to express $\omega$ in terms of $\lambda$. By varying the parameter $k$ (for a fixed value of the degeneracy parameter $\mu$), we can cover the whole diagram in parameter space. A complete description of this diagram has been given by Chavanis (2002c) in the non-rotating case.

\begin{figure}
\centering
\includegraphics[width=8.8cm]{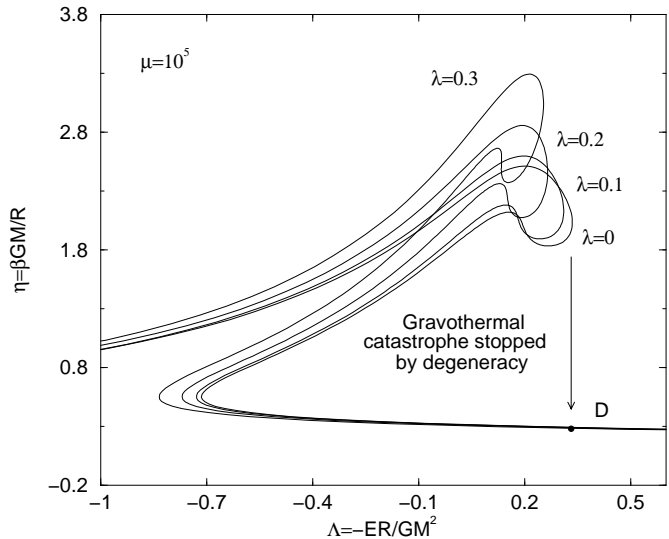}
\caption{Equilibrium phase diagram of self-gravitating fermions giving the inverse temperature $\eta$ as a function of minus the energy $\Lambda$ for different values of the angular momentum $\lambda$ and for a degeneracy parameter $\mu=10^{5}$.}
\label{FERMI_LeMu5}
\end{figure}

\begin{figure}
\centering
\includegraphics[width=8.8cm]{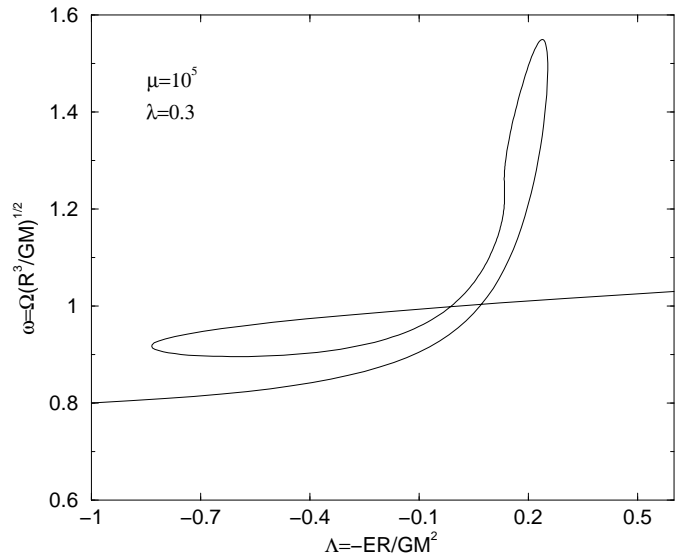}
\caption{Angular velocity vs energy for self-gravitating fermions with a degeneracy parameter $\mu=10^{5}$ and an angular momentum $\lambda=0.3$.}
\label{FERMI_LwMU5}
\end{figure}

In Fig. \ref{FERMI_LeMu5} we represent the equilibrium phase diagram of self-gravitating fermions for a degeneracy parameter $\mu=10^{5}$ and for different values of angular momentum. We observe that degeneracy has the effect of unwinding the spiral of Fig. \ref{LambdaetaMICRO}. For sufficiently large values of the degeneracy parameter, there is still gravitational collapse at $\Lambda_{c}$ accompanied by a rise of temperature, but this ``gravothermal catastrophe'' stops when the core of the system becomes degenerate. This leads to the formation of a ``fermion ball'' which contains a moderately large fraction of mass $\alpha M$ (at point $D$, we typically have $\alpha\simeq 0.2$). In the microcanonical ensemble, the decrease of potential energy in the core is compensated by an increase of temperature. Therefore, the mass $(1-\alpha)M$ contained in the halo undergoes an expansion which, in our model, is arrested by the walls of the box. As a result, the density of the halo is almost uniform. Typical density profiles are given by Chavanis \& Sommeria (1998) in the non rotating case. The expansion of the halo explains why the moment of inertia of the system increases despite the formation of a massive nucleus. Therefore, the angular velocity {\it decreases} during the collapse contrary to what might be expected. The angular velocity is represented as a function of energy in Fig. \ref{FERMI_LwMU5}. It has a complicated behaviour which corresponds to the unwindement of the spiral of Fig. \ref{Lomega}.

\begin{figure}
\centering
\includegraphics[width=8.8cm]{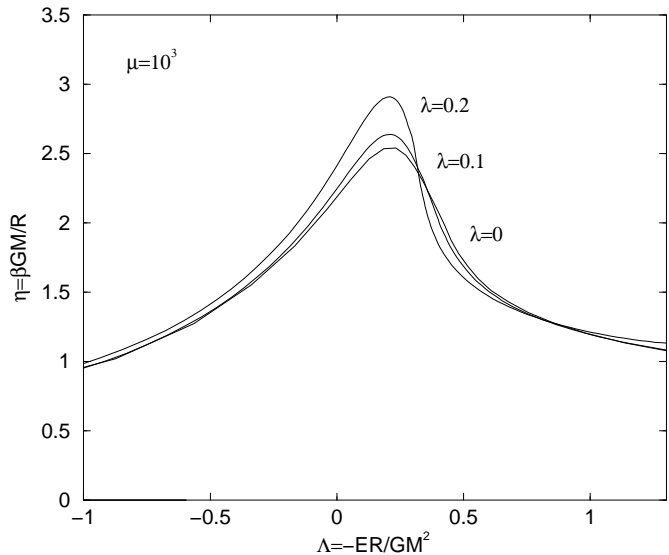}
\caption{Same as Fig. \ref{FERMI_LeMu5} for $\mu=10^{3}$. For high degeneracy, the spiral is unwound and the gravitational phase transition is suppressed in the microcanonical ensemble.}
\label{FERMI_LeMU3}
\end{figure}

\begin{figure}
\centering
\includegraphics[width=8.8cm]{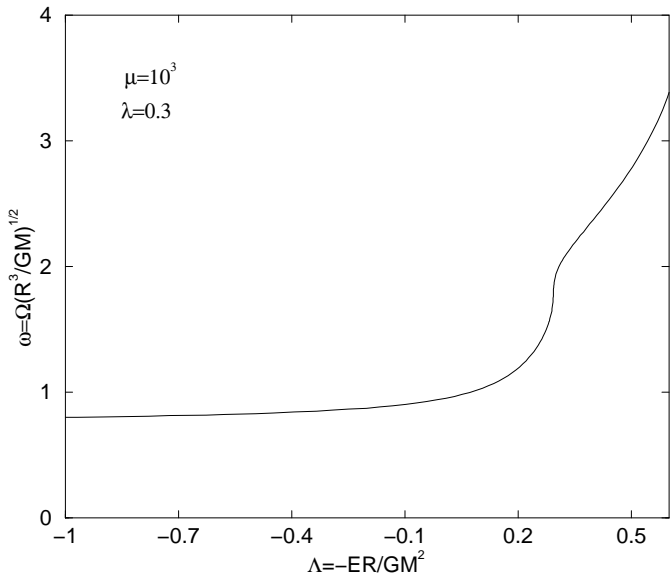}
\caption{Same as Fig. \ref{FERMI_LwMU5} for $\mu=10^{3}$. }
\label{FERMI_LomegaMU3}
\end{figure}

For smaller values of the degeneracy parameter, the gravitational phase transition is suppressed (Chavanis \& Sommeria 1998, Chavanis 2002c) and the equilibrium phase diagram has the structure of Fig. \ref{FERMI_LeMU3}. This diagram is similar to the one found by Fliegans \& Gross (2002) in their two-dimensional model of rotating self-gravitating systems. Had they used a smaller value of their cut-off radius (which plays the role of the inverse of our degeneracy parameter), they would have probably obtained a diagram similar to the one of Fig. \ref{FERMI_LeMu5}. However, when the cut-off radius (or degeneracy) is sufficiently large, the spiral unwinds and the $\eta$-$\Lambda$ curve is univalued like in Fig. \ref{FERMI_LeMU3}. For low energies, the equilibrium states have a core-halo structure with a partially degenerate nucleus and a dilute envelope. As energy decreases any further, the nucleus contains more and more mass and becomes smaller and smaller (a property of the $R\sim M^{-1/3}$ law of degenerate configurations). For rotating systems described in the microcanonical ensemble (fixed $E$ and $L$), the resulting decrease of moment of inertia is accompanied by an increase of angular velocity as shown in Fig. \ref{FERMI_LomegaMU3}. In this diagram, the $\omega-\Lambda$ spiral of Fig. \ref{Lomega} is completely unwound and the angular velocity increases monotonically with $\Lambda$.

\begin{figure}
\centering
\includegraphics[width=8.8cm]{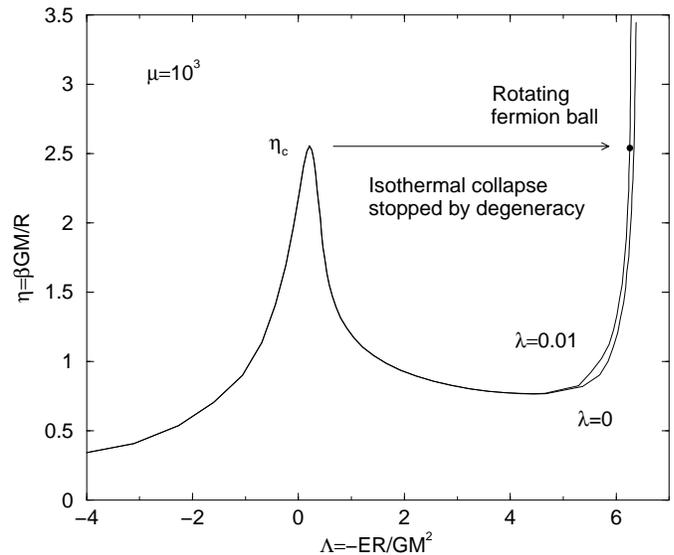}
\caption{Equilibrium phase diagram for self-gravitating fermions with a degeneracy parameter $\mu=10^{3}$ and an angular momentum $\lambda=0.01$. For $T<T_{c}$, i.e., $\eta>\eta_{c}$, the system undergoes an isothermal collapse leading to a rotating fermion ball containg a large fraction of mass and angular momentum.}
\label{FERMI_LetaMA0p01}
\end{figure}

\begin{figure}
\centering
\includegraphics[width=8.8cm]{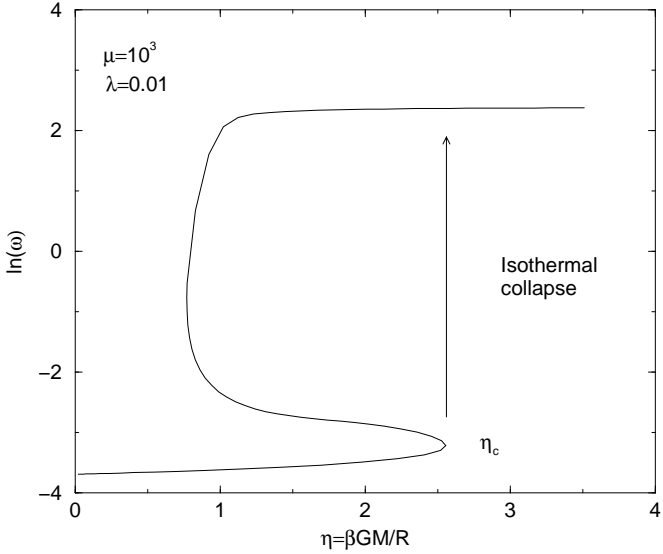}
\caption{Angular velocity vs temperature plot for $\mu=10^{3}$ and $\lambda=0.01$. The gravitational collapse below $T_{c}$ is accompanied by a discontinuous jump of angular velocity due to the moment of inertia decrease. The resulting fermion ball rotates with angular velocity $\Omega\simeq 0.1891 G^{1/2}M^{1/2}R_{0}^{-3/2}$ (see Appendix \ref{sec_app1}) corresponding to an increase $\sim 270$ of the initial angular velocity. The dimensionless rotation rate $v\simeq 3.98\ 10^{-3}$ is sufficiently small to justify the perturbative approach followed in that paper.}
\label{Fermi_etaomega0p01}
\end{figure}

In Fig. \ref{FERMI_LetaMA0p01} we have focused our attention to what happens close to the minimum energy.  For $E\rightarrow E_{min}$, $T\rightarrow 0$ and the system has the same structure as a cold white dwarf star (see Appendix \ref{sec_app1}). For our perturbative analysis to be valid, we have taken a very small angular velocity $\lambda=0.01$.  We shall describe the diagram of Fig. \ref{FERMI_LetaMA0p01} in the $(L,T)$ ensemble in which the angular momentum and the temperature are assumed given (this is a situation intermediate between microcanonical and canonical ensembles). For $T>T_{c}$, the system is in a gaseous phase with a smoothly decreasing density profile. For $T<T_{c}$, the system undergoes an isothermal collapse that only stops when gravity is balanced by the degeneracy pressure. The result of this phase transition is a ``fermion ball'' which contains almost {all} the mass, unlike in the microcanonical ensemble at the point of gravothermal catastrophe.  For rotating systems, this isothermal collapse is accompanied by a discontinuous rise of angular velocity at the critical temperature $T_{c}$ (see Fig. \ref{Fermi_etaomega0p01}). Therefore, even if the initial rotation of the system is negligible in the gaseous phase, after collapse the ``fermion ball'' can have appreciable rotation as suggested in Fig. \ref{Fermi_etaomega0p01}. Its structure is then similar to a distorted polytrope of index $n=3/2$ as  computed by Chandrasekhar (1933) in the limit of slow rotation.

\section{Conclusion}
\label{sec_conclusion}

In this paper, we have considered the effect of a small rotation on
the thermodynamical stability of self-gravitating systems. We have
worked in a finite box in order to make a clear connexion with the
Antonov problem for non-rotating systems and render the statistical
mechanics of these objects rigorous. Physically, this idealization
means that our isothermal system is surrounded by a medium which fixes
its size. We have found that for rotating systems, the well-known
inequivalence of statistical ensembles for self-gravitating systems
manifests itself in a striking manner: the instability is advanced in
the microcanonical ensemble and delayed in the canonical one. In
addition, we have found a connexion between the onset of instability
and the configuration of maximum flattening in the series of
equilibria.  These results have be generalized to the case of
self-gravitating fermions.

On the other hand, the case of rapidly rotating isothermal
configurations is interesting because new, non trivial, structures can
emerge as maximum entropy states (i.e., most probable states). The
classification of such structures is complicated because many
bifurcations can occur depending on the values of the control
parameters $(M,E,L)$. In particular, we must be careful to select only
entropy {\it maxima}, discarding the critical points of entropy which
are only saddle points. This can be obtained either by solving the
Boltzmann-Poisson equation (Votyakov et al. 2002) and checking the
stability of the solutions or by using relaxation equations towards
the maximum entropy state (Chavanis et al. 1996). These studies are
important in order to obtain a classification of the most probable
configurations of self-gravitating systems. The second approach
(relaxation methods) will be explored in a future paper (Rosier \&
Chavanis, in preparation).

\begin{acknowledgements}

I acknowledge interesting discussions with J. Katz. I have also benefited from stimulating interactions with the participants of the Les Houches School of Physics on ``Dynamics and thermodynamics of systems with long range interactions'' (February 2002).

\end{acknowledgements}

\appendix

\section{The entropy of a rotating self-gravitating Fermi gas}
\label{sec_app2}

The Fermi-Dirac entropy (\ref{fermions1bis}) can be written in the equivalent form 
\begin{eqnarray}
S=-\int\biggl\lbrace {f\over\eta_{0}}\ln\biggl ({f/\eta_{0}\over 1-f/\eta_{0}}\biggr )+\ln\biggl (1-{f\over \eta_{0}}\biggr )\biggr\rbrace d^{3}{\bf r}d^{3}{\bf v}.\nonumber\\
\label{B1}
\end{eqnarray} 
Inserting the distribution function 
\begin{eqnarray}
f={\eta_{0}\over 1+k e^{\Psi}e^{\beta{w^{2}\over 2}}},
\label{B2}
\end{eqnarray}
in Eq. (\ref{B1}), we obtain
\begin{eqnarray}
\eta_{0}S=M\ln k+\int \rho\Psi d^{3}{\bf r}+\beta K_{th}\nonumber\\
+\eta_{0}\int \ln\biggl (1+{1\over k}e^{-\Psi}e^{-\beta {w^{2}\over 2}}\biggr )d^{3}{\bf r}d^{3}{\bf v}.
\label{B3}
\end{eqnarray}
The last integral can be reduced by an integration by parts to the expression
$(2/3)\beta K_{th}$. Therefore,
\begin{eqnarray}
\eta_{0}S=M\ln k+\int \rho\Psi d^{3}{\bf r}+{5\over 3}\beta K_{th}.
\label{B4}
\end{eqnarray}
Now, using the definition of $\Psi$, we get
\begin{eqnarray}
\eta_{0}S=M\ln k+2\beta W-\beta K_{rot}-M\beta\Phi_{0}+{5\over 3}\beta K_{th}.
\label{B5}
\end{eqnarray}
Since $W=E-K$ and $K_{th}=K-K_{rot}$, the foregoing expression can be rewritten
\begin{eqnarray}
\eta_{0}S=M\ln k+2\beta E-{1\over 3}\beta K-M\beta\Phi_{0}-{8\over 3}\beta K_{rot},
\label{B6}
\end{eqnarray}
or, according to Eqs. (\ref{energy4bis}) and (\ref{energy6}),
\begin{eqnarray}
\eta_{0}S=M\ln k+{7\over 3}\beta E-{1\over 3}\beta \oint p{\bf r}d{\bf S}-M\beta\Phi_{0}-{4\over 3}\beta {\bf L}\cdot {\bf\Omega}.\nonumber\\
\label{B7}
\end{eqnarray}
The pressure at the surface of the sphere can be calculated with Eqs. (\ref{energy2bis}) and (\ref{B2}). After simplification, we get
\begin{eqnarray}
{\eta_{0}S\over M}=\ln k-{7\over 3}\eta \Lambda -\beta\Phi_{0}-{4\over 3}\eta\lambda\omega\nonumber\\
-{2\over 9}\eta {\alpha^{10}\over\mu^{4}}\biggl\lbrace {I_{3/2}(ke^{\psi(\alpha)})\over I_{1/2}(k)^{5}}-{3\over 2}v{I_{1/2}(ke^{\psi(\alpha)})\over I_{1/2}(k)^{5}}\phi_{0}(\alpha)\biggr\rbrace.
\label{B8}
\end{eqnarray}
Using Eq. (\ref{entropy8}), which remains valid for self-gravitating fermions,  and Eq. (\ref{fermions9}) we finally arrive at the form (\ref{fermions13}).

\section{The energy of a rotating polytrope}
\label{sec_app1}

In this Appendix, we derive a simple analytic formula for the potential energy  of a slowly rotating polytrope of index $n$. The index $n=3/2$ describes a completely degenerate Fermi gas at zero temperature, which is a particular limit of the model studied in Sec. \ref{sec_fermions}. 

For a self-gravitating system rotating with constant angular velocity $\Omega$, the condition of hydrostatic equilibrium in the rotating frame can be written 
\begin{eqnarray}
\nabla p=-\rho\nabla\Phi_{eff},
\label{A1}
\end{eqnarray}
where $\Phi_{eff}$ is the effective potential defined in Eq. (\ref{mf10a}). Now, for a polytropic gas,
\begin{eqnarray}
p=K\rho^{1+{1\over n}},
\label{A2}
\end{eqnarray}
and the equation of hydrostatic equilibrium can be rewritten as 
\begin{eqnarray}
(n+1)\nabla \biggl ({p\over\rho}\biggr )=-\nabla\Phi_{eff}.
\label{A3}
\end{eqnarray}
This equation integrates to give
\begin{eqnarray}
(n+1)\ p=\rho \bigl (\Phi_{eff}^{b}-\Phi+{1\over 2}|{\bf \Omega}\times {\bf r}|^{2}\bigr ),
\label{A4}
\end{eqnarray}
where $\Phi_{eff}^{b}$ is the effective potential at the surface of the polytrope. On integrating Eq. (\ref{A4}) over the volume of the configuration, we obtain
\begin{eqnarray}
(n+1)\int p \ d^{3}{\bf r}=M\Phi_{eff}-2W+{1\over 2}{\bf L}\cdot {\bf\Omega}.
\label{A5}
\end{eqnarray}
Inserting this last relation in Eq. (\ref{energy3}), and recalling that the pressure vanishes on the surface of a polytrope, we get
\begin{eqnarray}
W={1\over 5-n}\biggl\lbrack 3M\Phi_{eff}^{b}+\biggl (n+{5\over 2}\biggr ){\bf L}\cdot {\bf\Omega}\biggr \rbrack.
\label{A6}
\end{eqnarray}
This general expression was derived by Chandrasekhar (1961) in a slightly different manner. We now consider the case of slowly rotating polytropes. In that case, an explicit expression for $\Phi_{eff}^{b}$ can be deduced from the study of Chandrasekhar (1933) on distorted polytropes. The density profile can be written
\begin{eqnarray}
\rho=\rho_{0}\Theta^{n},
\label{A7}
\end{eqnarray}
with
\begin{eqnarray}
\Theta=\theta+v\biggl \lbrace \psi_{0}(\xi)+\sum_{j=1}^{+\infty}A_{j}\psi_{j}(\xi)P_{j}(\mu)\biggr\rbrace.
\label{A8}
\end{eqnarray}
The quantities have their usual meaning: $\rho_{0}$ is the central density, $v=\Omega^{2}/2\pi G\rho_{0}$, $\theta$ is the solution of the Lane-Emden equation of index $n$ and $P_{j}(\mu)$ are Legendre polynomials. The functions $\psi_{j}(\xi)$ have been defined and tabulated by Chandrasekhar (1933). The inner and outer potentials are given by
\begin{eqnarray}
\Phi_{int}=\Phi_{eff}^{b}-R\biggl\lbrack \Theta-{1\over 6}v\xi^{2}
(1-P_{2}(\mu))\biggr\rbrack,
\label{A9}
\end{eqnarray} 
\begin{eqnarray}
\Phi_{ext}=-R\biggl\lbrack {C_{0}\over\xi}+v\sum_{j=1}^{+\infty}{C_{j}\over \xi^{j+1}}P_{j}(\mu)\biggr\rbrack,
\label{A10}
\end{eqnarray} 
with $R=(n+1)K\rho_{0}^{1/n}$. The constant $A_{j}$, $C_{j}$ are determined by requiring the continuity of $\Phi$ and its derivative on a sphere of radius $\xi_{1}$, the first zero of the Emden's function with index $n$. For our purposes, we need to consider the equations obtained for $j=0$:
\begin{eqnarray}
R{C_{0}\over \xi_{1}}=Rv\psi_{0}(\xi_{1})-{1\over 6}Rv\xi_{1}^{2}-\Phi_{eff}^{b},
\label{A11}
\end{eqnarray} 
\begin{eqnarray}
-R{C_{0}\over \xi_{1}^{2}}=R\theta_{1}'+Rv\psi_{0}'(\xi_{1})-{1\over 3}Rv\xi_{1}.
\label{A12}
\end{eqnarray} 
In Eq. (\ref{A11}), we have used $\theta(\xi_{1})=0$. Now, the relation between the mass and the central density is given by (Chandrasekhar 1933, Eq. (40)):
\begin{eqnarray}
M=-4\pi\biggl\lbrack {(n+1)K\over 4\pi G}\rho_{0}^{3-n\over 3n}\biggr\rbrack^{3/2}\xi_{1}^{2}\theta'_{1}\biggl\lbrack 1+v{{1\over 3}\xi_{1}-\psi'_{0}(\xi_{1})\over |\theta'_{1}|}\biggr\rbrack.\nonumber\\
\label{A13}
\end{eqnarray} 
On the other hand, we define the characteristic radius $R_{\Omega}$ of a rotating polytrope by the relation (see Chandrasekhar 1933, Eq. (9)): 
\begin{eqnarray}
R_{\Omega}=\biggl\lbrack {(n+1)K\over 4\pi G}\rho_{0}^{{1\over n}-1}\biggr\rbrack^{1/2}\xi_{1}.
\label{A14}
\end{eqnarray} 
Inserting Eq. (\ref{A12}) in Eq. (\ref{A13}) and using Eq. (\ref{A14}), we obtain
\begin{eqnarray}
R{C_{0}\over \xi_{1}}={GM\over R_{\Omega}}.
\label{A15}
\end{eqnarray}
On the other hand, we can check that
\begin{eqnarray}
Rv={2\Omega^{2}\over \xi_{1}^{2}}R_{\Omega}^{2}.
\label{A16}
\end{eqnarray}
Substituting these results in Eq. (\ref{A11}), we find that the effective potential at the surface of a slowly rotating polytrope is given by
\begin{eqnarray}
\Phi_{eff}^{b}=-{GM\over R_{\Omega}}+2\Omega^{2}R_{\Omega}^{2}\biggl ({\psi_{0}(\xi_{1})\over\xi_{1}^{2}}-{1\over 6}\biggr ).
\label{A17}
\end{eqnarray} 

We now need to relate the characteristic radius $R_{\Omega}$ to the total mass $M$ of the system. Eliminating the central density $\rho_{0}$ between Eq. (\ref{A13}) and Eq. (\ref{A14}), and expressing $v$ in terms of $R_{\Omega}$ with the aid of Eq. (\ref{A16}), we obtain to order $\Omega^{2}$:
\begin{eqnarray}
R_{\Omega}=R_{0}\biggl\lbrack 1+\Omega^{2}{S_{n}K^{3n\over 3-n}\over 2\pi G^{2n+3\over 3-n}M^{2n\over 3-n}}\biggr\rbrack,
\label{A18}
\end{eqnarray} 
where $R_{0}$ is the radius of a non-rotating polytrope. It is related to its mass $M$ by the relation
\begin{eqnarray}
GN_{n}M^{n-1\over n}R_{0}^{3-n\over n}=K.
\label{A19}
\end{eqnarray} 
The constant $N_{n}$ is defined and tabulated in Chandrasekhar (1942) for different values of the polytropic index $n$. The constant $S_{n}$ appearing in Eq. (\ref{A18}) is given by
\begin{eqnarray}
S_{n}={n-1\over 3-n}\biggl ({n+1\over 4\pi}\biggr )^{n\over 1-n}N_{n}^{2n^{2}\over (1-n)(3-n)}\xi_{1}^{2n\over 1-n}{{1\over 3}\xi_{1}-\psi_{0}'(\xi_{1})\over |\theta'_{1}|},\nonumber\\
\label{A20}
\end{eqnarray}  
and typical values are listed in Table \ref{table}. 

From the above results, we can obtain an explicit expression for the potential energy of a rotating polytrope. For an axisymmetrical system, we have the relation ${\bf L}=I {\bf \Omega}$ where $I$ is the axial moment of inertia (\ref{angular3}). To our order of approximation, we just need to determine the value of $I$ for a non-rotating polytrope. Using the relations $\rho=\rho_{0}\theta^{n}$ and $r=\xi R_{0}/\xi_{1}$, expressing the central density as a function of $R_{0}$ by the relation (\ref{A14}) and using the mass-radius relation (\ref{A19}), we obtain
\begin{eqnarray}
I={\cal I}_{n}MR_{0}^{2},
\label{A21}
\end{eqnarray}    
with
\begin{eqnarray}
{\cal I}_{n}={8\pi\over 3}\biggl\lbrack {4\pi\over (n+1)N_{n}}\biggr \rbrack^{n\over 1-n}\xi_{1}^{5-3n\over n-1}\int_{0}^{\xi_{1}}\theta^{n}\xi^{4}d\xi.
\label{A22}
\end{eqnarray}
Therefore, according to Eqs. (\ref{A6}) (\ref{A17}) and (\ref{A21}), the potential energy of a slowly rotating polytrope of index $n$ is given by
\begin{eqnarray}
W={1\over 5-n}\biggl( -{3GM^{2}\over R_{\Omega}}+Q_{n}MR_{0}^{2}\Omega^{2}\biggr ),
\label{A23}
\end{eqnarray}  
with
\begin{eqnarray}
Q_{n}=6{\psi_{0}(\xi_{1})\over\xi_{1}^{2}}-1+\biggl (n+{5\over 2}\biggr ){\cal I}_{n},
\label{A24}
\end{eqnarray} 
where $R_{0}$ and $R_{\Omega}$ are given by Eqs. (\ref{A19}) and (\ref{A18}) respectively. For $\Omega=0$, Eq. (\ref{A23}) reduces to the well-known Ritter's formula (see Chandrasekhar 1942). Note also that the dimensionless rotation parameter can be written
\begin{eqnarray}
v=Z_{n}{\Omega^{2}R_{0}^{3}\over GM},
\label{A24a}
\end{eqnarray} 
with
\begin{eqnarray}
Z_{n}={1\over 2\pi}\biggl ({n+1\over 4\pi}\biggr )^{n\over 1-n}N_{n}^{n\over 1-n}\xi_{1}^{2n\over 1-n}.
\label{A24b}
\end{eqnarray} 

The polytropic indices $n=1$ and $n=3$ are special. For $n=1$, the problem can be solved analytically. The following results are well-known (Chandrasekhar 1942, 1933): $\xi_{1}=\pi$, $\theta'_{1}=-1/\pi$, $\psi_{0}(\xi_{1})=1$, $\psi'_{0}(\xi_{1})=1/\pi$, $N_{1}=2/\pi$. According to Eq. (\ref{A15}), the radius $R_{\Omega}=R_{0}=(K/2\pi G)^{1/2}\xi_{1}$ is independant of the central density and is uniquely determined by the value of $K$. The mass $M$ is arbitrary and independant of the radius. Eqs. (\ref{A21}) (\ref{A24a}) remain valid with ${\cal I}_{1}=2(\pi^{2}-6)/3\pi^{3}$ and $Z_{1}=2/\pi^{2}$. The other quantities can be obtained from these results (see Table \ref{table}). The case $n=3$ corresponds to the reverse situation. For $\Omega=0$, the mass is given by $M_{0}=(K/GN_{3})^{3/2}$ and the radius is arbitrary. For $\Omega\neq 0$, we can still consider that the radius $R_{\Omega}=R_{0}$ is a free parameter and determine the mass according to Eq. (\ref{A13}), which can be written  
\begin{eqnarray}
M=M_{0}\biggl\lbrack 1+\Omega^{2}S'_{3}{G^{1/2}R_{\Omega}^{3}\over 2K^{3/2}}\biggr\rbrack,
\label{MM0}
\end{eqnarray} 
with
\begin{eqnarray}
S'_{3}={\pi^{1/2}\over\xi_{1}^{3}}{{1\over 3}\xi_{1}-\psi_{0}'(\xi_{1})\over |\theta'_{1}|}\simeq 0.03309.
\label{MM0bis}
\end{eqnarray}

 \begin{table}
\caption[]{The constants of slowly rotating polytropes.}
         \label{table}
      \[
         \begin{array}{l l l l l l l l l l  }
            \hline
            \noalign{\smallskip}
  {n} & N_{n}  & S_{n} & {\cal I}_{n} & Q_{n} & Z_{n} \\
            \noalign{\smallskip}
            \hline
            \noalign{\smallskip}
      {1} & 0.63662 & 0 & 0.08320 & -0.1009 & 0.20264\\
                  
      {1.5}   & 0.42422 & 8.73193 & 0.20461 & 0.40008 & 0.11127\\ 
                 
      {2}  &  0.36475 & 557.60 & 0.15485 & 0.30331 & 0.05846\\ 
 
      {3}  &  0.36394 & \infty & 0.07536 & 0.15087 & 0.01230\\ 	

      {4}  & 0.47720 & -3.8\ 10^{-5} & 0.02257 & 0.04434 & 0.00107\\

           \noalign{\smallskip}
            \hline
         \end{array}
     \]
   \end{table}

We now consider the case of a degenerate Fermi gas at zero temperature. As is well-known, this system is equivalent to a polytrope of index $n=3/2$. In addition, the constant $K$ is explicitly given by
\begin{eqnarray}
K={1\over 5}\biggl ({3\over 4\pi\eta_{0}}\biggr )^{2/3},
\label{A25}
\end{eqnarray} 
where $\eta_{0}$ is the distribution function of a completely degenerate Fermi gas. According to the Virial theorem (\ref{energy6}), the total energy $E=K+W$ of this system is $E={W\over 2}$. Therefore, using Eq. (\ref{A23}), we have
\begin{eqnarray}
E=-{3GM^{2}\over 7R_{\Omega}}+{1\over 7}Q_{3/2}MR_{0}^{2}\Omega^{2}.
\label{A26}
\end{eqnarray} 
According to Eq. (\ref{A19}), the mass-radius relation of a non rotating ``fermion ball'' is
\begin{eqnarray}
MR_{0}^{3}={\chi\over \eta_{0}^{2}G^{3}},
\label{A27}
\end{eqnarray}
with
\begin{eqnarray}
\chi\equiv {1\over 125}\biggl ({3\over 4\pi}\biggr )^{2}{1\over N_{3/2}^{3}}\simeq 5.9723\ 10^{-3}.
\label{A27bis}
\end{eqnarray}
For a rotating fermion ball we have, according to Eq.  (\ref{A18}),
\begin{eqnarray}
R_{\Omega}=R_{0}\biggl (1+{\kappa\Omega^{2}\over 2\pi G^{4}M^{2}\eta_{0}^{2}}\biggr ),
\label{A28}
\end{eqnarray} 
with
\begin{eqnarray}
\kappa\equiv {1\over 125}\biggl ({3\over 4\pi}\biggr )^{2}S_{3/2}\simeq 3.9813\ 10^{-3}.
\label{A28bis}
\end{eqnarray} 
Then, the energy (\ref{A26}) can be expressed in terms of $M$, $G$, $\eta_{0}$ and $\Omega$. In order to make the link with the variables introduced in Sec. \ref{sec_fermions}, which are normalized by the box radius $R$, we note the relation
\begin{eqnarray}
{R_{0}\over R}=(512\pi^{4}\chi)^{1/3}\mu^{-2/3}={6.6784\over \mu^{2/3}},
\label{A29}
\end{eqnarray}   
which directly results from Eq. (\ref{A27}) and the definition of the degeneracy parameter $\mu$.

\end{document}